\documentclass[fleqn,usenatbib]{mnras}

\usepackage{newtxtext,newtxmath}

\usepackage[T1]{fontenc}
\usepackage{ae,aecompl}

\usepackage{graphicx}	
\usepackage{amsmath}	
\usepackage{amssymb}	
\usepackage{multirow}
\usepackage{color}
\definecolor{yaleblue}{rgb}{0.06, 0.3, 0.57}
\definecolor{myorange}{RGB}{245,156,74}
\definecolor{navy}{RGB}{25,25,112}
\hypersetup{draft}

\def\kms{km\,s$^{-1}$}
\def\Ha{H{$\alpha$}}
\def\Hb{H{$\beta$}}

\def\Hd{H{$\delta$}}
\def\ni{$^{56}$Ni}
\def\co{$^{56}$Co}
\def\fe{$^{56}$Fe}

\def\M{M$_{\odot}$}

\def\hx{SN2013hx}
\def\br{PS15br}
\def\sl1{SLSNe~I}
\def\es{SN2008es}
\def\lesssim{\mathrel{\hbox{\rlap{\hbox{\lower4pt\hbox{$\sim$}}}\hbox{$<$}}}}
\def\gtrsim{\mathrel{\hbox{\rlap{\hbox{\lower4pt\hbox{$\sim$}}}\hbox{$>$}}}}
\newcommand{\gps}{\ensuremath{g_{\rm P1}}}
\newcommand{\rps}{\ensuremath{r_{\rm P1}}}
\newcommand{\ips}{\ensuremath{i_{\rm P1}}}
\newcommand{\zps}{\ensuremath{z_{\rm P1}}}

\title[SLSNe~II]{On the nature of Hydrogen-rich Superluminous Supernovae}

\author[C. Inserra]{C. Inserra,$^{1,2}$\thanks{E-mail: C.Inserra@soton.ac.uk}
S. J. Smartt$^{2}$,
E. E. E. Gall$^{2,3}$,
G. Leloudas$^{4}$,
T.-W. Chen$^{5}$,
\newauthor S. Schulze$^{6,7}$,
A. Jerkstrand$^{8}$,
M. Nicholl$^{9}$,
J. P. Anderson$^{10}$,
I. Arcavi$^{11,12,13}$,
\newauthor S. Benetti$^{14}$,
R. A. Cartier$^{1}$,
M. Childress$^{1}$,
M. Della Valle$^{15}$,
H. Flewelling$^{16}$,
\newauthor M. Fraser$^{17}$,
A. Gal-Yam$^{18}$,
C. P. Guti\'errez$^{1}$,
G. Hosseinzadeh$^{11,12}$,
D. A. Howell$^{11,12}$,
\newauthor M. Huber$^{16}$,
E. Kankare$^{2}$,
T. Kr\"{u}hler$^{5}$,
E. A. Magnier$^{16}$,
K. Maguire$^{1}$,
\newauthor C. McCully$^{11,12}$,
S. Prajs$^{1}$,
N. Primak$^{16}$,
R. Scalzo$^{19,20}$,
B. P. Schmidt$^{21}$, M. Smith$^{1}$,
\newauthor K. W. Smith$^{2}$,
B. E. Tucker$^{19,20}$,
S. Valenti$^{22}$,
M. Wilman$^{16}$,
D. R. Young$^{2}$, F. Yuan$^{19,20}$
\\
$^1$ Department of Physics and Astronomy, University of Southampton, Southampton, SO17 1BJ, UK\\
$^2$ Astrophysics Research Centre, School of Mathematics and Physics, Queens University Belfast, Belfast BT7 1NN, UK\\
$^3$ Max-Planck-Institut f\"ur Astrophysik, Karl-Schwarzschild-Str. 1, DE-85748 Garching-bei-M\"{u}nchen, Germany\\
$^4$Dark Cosmology Centre, Niels Bohr Institute, University of Copenhagen, Juliane Maries vej 30, 2100 Copenhagen, Denmark\\
$^5$Max-Planck-Institut f{\"u}r Extraterrestrische Physik, Giessenbachstra\ss e 1, 85748, Garching, Germany\\
$^6$Instituto de Astrof\'isica, Facultad de F\'isica, Pontificia Universidad Cat\'olica de Chile, Vicu\~{n}a Mackenna 4860, 7820436 Macul, Santiago, Chile\\
$^7$Millennium Institute of Astrophysics, Vicu\~{n}a Mackenna 4860, 7820436 Macul, Santiago, Chile\\
$^8$Max-Planck Institut f{\"u}r Astrophysik, Karl-Schwarzschild-Str. 1, D-85748 Garching, Munich, Germany\\
$^9$Harvard-Smithsonian Center for Astrophysics, 60 Garden Street, Cambridge, Massachusetts 02138, USA\\
$^{10}$European Southern Observatory, Alonso de Córdova 3107, Casilla 19, Santiago, Chile\\
$^{11}$Las Cumbres Observatory Global Telescope Network, 6740 Cortona Dr., Suite 102 Goleta, Ca 93117\\
$^{12}$Department of Physics, University of California, Santa Barbara, CA 93106-9530, USA\\
$^{13}$Einstein fellow\\
$^{14}$INAF, Osservatorio Astronomico di Padova, vicolo dell'Osservatorio 5, 35122, Padova, Italy\\
$^{15}$INAF Osservatorio Astronomico di Capodimonte, Salita Moiariello 16, I-80131 Napoli, Italy\\
$^{16}$Institute for Astronomy, University of Hawaii at Manoa,Honolulu, HI 96822, USA\\
$^{17}$School of Physics, O'Brien Centre for Science North, University College Dublin, Belfield, Dublin 4, Ireland\\
$^{18}$Department of Particle Physics and Astrophysics, Weizmann Institute of Science, Rehovot 7610001, Israel\\
$^{19}$Research School of Astronomy and Astrophysics, Australian National University, Canberra, ACT 2611, Australia\\
$^{20}$ARC Centre of Excellence for All-Sky Astrophysics (CAASTRO), The Australian National University, Canberra, ACT 2611, Australia\\
$^{21}$Australian National University, Canberra, ACT 2611, Australia\\
$^{22}$Department of Physics, University of California, Davis, CA 95616, USA\\
}

\date{Accepted XXX. Received YYY; in original form ZZZ}

\pubyear{}

\begin{document}
\label{firstpage}
\pagerange{\pageref{firstpage}--\pageref{lastpage}}
\maketitle

\begin{abstract}
We present two hydrogen-rich superluminous supernovae (SLSNe), namely \hx\/ and \br. These objects, together with \es\/ are the only SLSNe showing a distinct, broad \Ha\/ feature during the photospheric phase and also do not show any sign of strong interaction between fast moving ejecta and circumstellar shells in their early spectra. Despite \br\/ peak luminosity is fainter than the other two objects, the spectrophotometric evolution is similar to \hx\/ and different than any other supernova in a similar luminosity space. We group all of them as SLSNe II and hence distinct from the known class of SLSN IIn. Both transients show a strong, multi-component H$\alpha$ emission after 200 days past maximum which we interpret as an indication of interaction of the ejecta with an asymmetric, clumpy circumstellar material. 
The spectra and photometric evolution of the two objects are similar to type II supernovae, although they have much higher luminosity and evolve on slower timescales. This is qualitatively similar to how SLSNe I compare with normal type Ic in that the former are brighter and evolve more slowly. 
We apply a magnetar and an interaction semi-analytical codes to fit the light curves of our two objects and \es. 
The overall observational dataset would tend to favour the magnetar, or central engine, model as the source of the peak luminosity although the clear signature of late-time interaction indicates that interaction can play a role in the luminosity evolution of SLSNe II at some phases.
\end{abstract}

\begin{keywords}
supernovae: general - supernovae: individual (\hx, \br, \es) - supernovae: interactions - stars: magnetars
\end{keywords}



\section{Introduction}\label{sec:intro}

Over the past few years, the current generation of wide-field optical surveys have discovered a new class of intrinsically bright transients. They show absolute magnitudes at maximum light of $M_{\rm AB} \sim-21$ mag, total radiated energies of  order
$10^{51}$ erg \citep[e.g.][]{qu11} and now go by the common name of superluminous supernovae 
\citep[SLSNe,][]{gy12}. They 
 are factors of 5 to 100  brighter than type Ia  or normal core-collapse supernovae and have recently been classified on their spectrophotometric behaviour rather than a simple magnitude threshold \citep[e.g.][]{
papa15,lu16,pr17}.

Two distinct groups have emerged so far. The first, and best studied, includes hydrogen-free SLSNe which have spectra  at maximum light  showing a blue continuum and a distinctive 
``W"-shaped spectra feature in absorption at $\sim4200$~\AA\/.  The absorption has
been identified as O\,{\sc ii}  and these SNe have been labelled SLSNe I 
\citep{qu11,gy12}. Their spectra at about 30 days after peak are very similar to normal or broadlined type Ic SNe at peak luminosity \citep{pasto10}, and hence are also called SLSNe Ic 
\citep{in13}. 

The second group is that of hydrogen-rich SLSNe. This includes the very bright, strongly interacting SNe such as SN2006gy \citep[e.g.][]{smi07,smm07,of07,ag09} - of which the enormous luminosity is mainly powered by the interaction of supernova ejecta with dense circumstellar (CSM) shells and, following the standard taxonomy, should be labeled  SLSNe IIn (due to clear hydrogen multicomponent and narrow emission lines exhibited since the early spectra). 

However, there is also one known SLSN which in many ways is similar to the type Ic SLSNe, but 
with a distinct and broad H$\alpha$ feature. This is \es\/, which was studied by 
 \cite{mi09} and \cite{ge09} and does not obviously show any signs of interaction between 
fast and slow moving ejecta and CSM shells. The favoured explanation for this object, which somewhat resembled normal luminosity type II SNe \citep{mi09}, was a core-collapse explosion of a non-standard progenitor star with a super wind and extended envelope \citep{ge09}. 
Recent papers have also shown SLSNe I with a weak, but distinct, 
multi-component \Ha\/ emission \citep{ben14} or with late-time \Ha\/ emission \citep{yan15,yan17}.  

It is plausible that there are multiple powering sources in any of these explosions. For example
a central engine may power the bulk of the luminosity, but then interaction between the 
magnetar powered ejecta and some CSM shells (of varying density) could provide additional 
energy output and alter the spectra morphology as observed for a nearby sample of slow SLSNe I \citep{in17}. 
Hereafter we will refer to events without obvious spectral signatures of interaction as SLSNe II whereas those with narrow spectral features, multicomponent profiles and obviously powered by interaction will be referred to as SLSNe IIn.
In this paper we present 
two more SLSNe II objects, followed in detail by the Public ESO Spectroscopic Survey for 
Transient Objects \citep[PESSTO;][]{pessto} with H$_0=72$ \kms\/,  $\Omega_{\rm M}=0.27$, $\Omega_{\lambda}=0.73$ adopted as standard cosmology.

\section{Sample}\label{sec:sample}
\subsection{\hx}\label{sec:hx}
\hx\/  (SMTJ013533283-5757506) was discovered by the SkyMapper Transient (SMT) and Supernova Survey \citep{keller07,atelscalzo13} on the 27th December 2013 in $g$ and $r$ bands. Three weeks before the discovery, on MJD 56632.55, a SMT image shows no detection of the transient to $r \simeq 20.28$ mag. Hence we can determine the epoch of explosion to around twenty days, at least as far as the sensitivity of the images allow. The object coordinates have been measured on our astrometrically calibrated images: $\alpha = 01^{\rm h}35^{\rm m}32^{\rm s}.83\pm0^{\rm s}.05$, $\delta = -57^{\rm o}57'50.6"\pm0".05$ (J2000). The object brightened slowly by $\sim1$ mag in the observed 
$g$-band from discovery to peak and the photometry is given in Table~\ref{table:sn13hx}. 
It was classified by the Public ESO Spectroscopic Survey of Transient Objects \citep[PESSTO,][]{pessto} around maximum as a hydrogen-rich superluminous supernova showing some similarities to SN2010gx \citep{pasto10} and CSS121015:004244+132827 \citep{ben14} on February 20 UT \citep{cam14,cbetscalzo14}. The spectrum showed a blue continuum with broad features in the blue together with a narrow emission feature at $\sim$7500~\AA\/ consistent with \Ha\/ setting the object at $z=0.125$ (see Table~\ref{table:sample} for \hx\/ main properties). It was then immediately selected by PESSTO as follow-up science targets and a combination of optical, near infrared (NIR) and ultraviolet (UV) photometric monitoring, together with optical spectroscopic monitoring, was carried out (see a summary in Table~\ref{table:obsshort}). Observations and data reduction are reported in Appendix~\ref{sec:data}.  
The Galactic reddening toward the position of the SN is E(B-V)~=~0.02 mag \citep{sf11}. The available spectra do not show Na~{\sc id} lines from the host galaxy, hence we adopt the Galactic reddening as total reddening.

Deep $gri$ images taken by PESSTO with the New Technology Telescope (NTT)+EFOSC2 on the 10th of December 2015, after the SN faded, find an extended, faint source at $\alpha = 01^{\rm h}35^{\rm m}32^{\rm s}.78\pm0^{\rm s}.05$, $\delta = -57^{\rm o}57'52.3"\pm0".05$ (J2000) which is at 2\arcsec\/ from SN location (4.36 kpc at $z=0.125$). The magnitudes of this galaxy are 
$g= 24.43\pm0.16$, $r=23.20\pm0.16$ and $i=21.82\pm0.16$ mag (Host A). 
This source was not detected in the NIR with deep $J,H,K$ images taken by PESSTO with the NTT+SOFI on the 17 December 2015
($J>23.1$ mag, see  Table~\ref{table:sn13hx} for the limits in each filter). 
Even deeper images from  Magellan+IMACS on the 1 February 2016 showed faint flux closer to the location 
of SN2013hx (within 0\farcs5) at magnitude $g= 24.71\pm0.38$, $r=24.55\pm0.35$ and $i=23.54\pm0.32$ mag (Host B).  This flux could either be the true host or residual flux from SN2013hx or a combination of both. Given that the Magellan detections are close to 3$\sigma$ significance and that star/galaxy separation is not reliable at these flux limits we avoid any conclusion on the true host \citep[but see][for an in depth analysis]{schulze17}. 

\begin{table*}
\caption{Main properties of the SLSNe II here presented.}
\begin{center}
\begin{tabular}{ccc}
\hline
\hline
&\hx & \br \\
\hline
Alternative &  SMTJ013533283-5757506  &CSS150226-112519+081418  \\
names&  & MLS150612-112519+081418\\
$\alpha$ (J2000.0)& $01^{\rm h}35^{\rm m}32^{\rm s}.83$ & $11^{\rm h}25^{\rm m}19^{\rm s}.22$  \\
$\delta$ (J2000.0)& $-57^{\rm o}57'50.6"$ &  $ 8^{\rm o}14'18.9"$ \\
$z$ & 0.125 & 0.101 \\
Peak $g$ (mag)& $-21.70$& $-20.22$\\
E(B-V) (mag) & 0.02 & 0.06 \\
L$_{\rm  ugriz}$ peak (x $10^{43}$ erg s$^{-1}$) & 10.78 & 2.43  \\
Light curve peak (MJD) & 56684.50 $\pm$ 1.00 & 57089.25 $\pm$ 2.00 \\
Host $r$ (mag)&$-15.58$(A),$-14.81$(B) &$-16.25$ \\
\hline
\end{tabular}
\end{center}
\label{table:sample}
\end{table*}%

\subsection{\br}\label{sec:br}
\br\/  was discovered by the Pan-STARRS Survey for Transients \citep[PSST,][]{atelhub15}\footnote{http://star.pst.qub.ac.uk/ps1threepi/psdb/} on  16 February  2015 at $w_{\rm P1}=19.10\pm0.02$ mag, confirmed as a transient source the day after and subsequently detected by the Catalina Real-time Transient Survey \citep[CRTS,][with IDs CSS150226-112519+081418 and MLS150612-112519+081418]{dr09} at $R=18.3$ mag on 26 February 2015. 
The closest pre-detection image is on the 16 January 2015 from CRTS images.
We measured the objects coordinates on our astrometrically calibrated images:  $\alpha = 11^{\rm h}25^{\rm m}19^{\rm s}.22\pm0^{\rm s}.05$, $\delta = 8^{\rm o}14'18.9"\pm0".05$ (J2000). The SN was observed to rise slowly with these multiple detections by PSST and CRTS and was then 
classified by PESSTO on the 11th of March 2015 \citep{atelfr15} as a superluminous supernova around maximum light at $z=0.101$ (see Table~\ref{table:sample} for \br\/ main properties).  
As for the previous target, also \br\/ was followed-up by the PESSTO consortium and monitored with optical+UV+NIR imaging and optical and NIR (only at late-time) spectroscopy, which revealed the presence of Balmer lines. A summary of \br\/ observations can be found in Table~\ref{table:obsshort}.
Thanks to the additional spectroscopy presented here we were able to secure the classification as SLSN II. The foreground reddening is E(B-V)~=~0.06 mag from \citet{sf11}. Also in this case the available spectra do not show Na~{\sc id} lines from the host galaxy and we only adopt the Galactic reddening.

We retrieved stacks from the Pan-STARRS1 Science Consortium  $3\pi$ survey 
\citep{to12,2012ApJ...756..158S,2013ApJS..205...20M} with total exposure 
times of  $g$ (720s), $r$ (590s), $i$ (1800s), $z$ (930s) from before the explosion of PS15br and which 
are unlikely to contain SN flux (see Appendix~\ref{sec:tab}, Table~\ref{table:sn15br}). Aperture photometry produced the following AB magnitudes for the extended host galaxy 
\gps$= 22.39\pm 0.11$, 
\rps$ = 22.18\pm 0.12$,
\ips$ = 21.70\pm 0.09$, 
\zps$ = 22.15\pm 0.11$ mag. 
We estimate the centre of this dwarf host galaxy 
is 0.6\arcsec\/ from SN location (which would be 1 kpc at $z=0.101$) and within the uncertainties is 
effectively coincident. 
The photometric measurements of the host are in agreement with the Sloan magnitudes of the galaxy SDSS J112519.21+081417.9 listed in the SDSS DR12 \citep{dr12}.

\begin{table*}
\caption{Observations log (see Appendix~\ref{sec:data} for further information).}
\begin{center}
\begin{tabular}{ccc}
\hline
\hline
Type & phase &\hx  \\
\hline
\multirow{ 2}{*}{Photometry} & early/photospheric ($\leq110$d) & Skymapper; NTT+EFOSC2; Swift+UVOT; LCO+Sinestro  \\
& late/nebular($> 110$d)  & NTT+EFOSC2; NTT+SOFI \\
\multirow{ 2}{*}{Spectroscopy} & early/photospheric ($\leq110$d) & NTT+EFOSC2; ANU+WiFeS;   \\
& late/nebular ($> 110$d) & NTT+EFOSC2; VLT+FORS2  \\
\hline
\hline
Type & phase &\br  \\
\hline
\multirow{ 2}{*}{Photometry} & early/photospheric ($\leq110$d) &  LCO+Sinestro; LT+IO:O; PSST; CSS; Swift+UVOT; NTT+SOFI  \\
& late/nebular($> 110$d)   & NTT+EFOSC2; LT+IO:O; NTT+SOFI\\
Polarimetry & early/photospheric ($\leq110$d) & VLT+FORS2 \\
\multirow{ 2}{*}{Spectroscopy} & early/photospheric ($\leq110$d) & NTT+EFOSC2; ANU+WiFeS; UH+SNIFS \\
& late/nebular ($> 110$d) & NTT+EFOSC2; VLT+XHOOTER\\
\hline
\end{tabular}
\end{center}
\label{table:obsshort}
\end{table*}%

\begin{figure*}
\includegraphics[width=18cm]{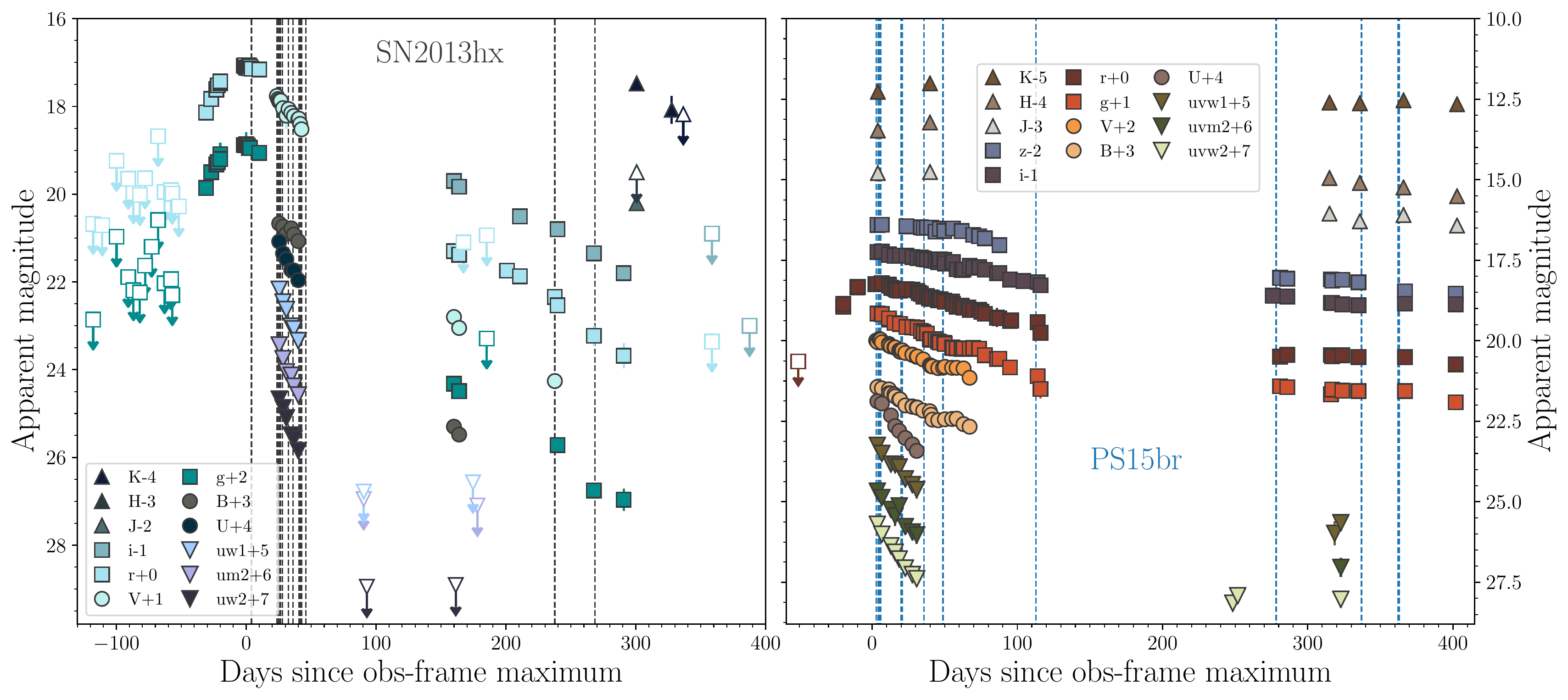}
\caption{Left: $uvw2,uvm2,uvw1,U,B,V,g,r,i,J,H,K$ light curve evolution of \hx\/ in the observed frame. Right: $uvw2,uvm2,uvw1,U,B,V,g,r,i,z,J,H,K$ light curve evolution of \br\/ in the observed frame. Open symbols denote limits. Phase is with respect to maximum
light at observed frame in $r$-band. Uncertainties are also reported The epochs of  \hx\/ and \br\/ spectra are marked with black and blue vertical dashed lines, respectively }
\label{fig:lc}
\end{figure*}

\section{Photometric, spectroscopic and polarimetric evolution}
\subsection{Light curves}\label{sec:lc}
\hx\/ shows a light curve with a 30 day rise in the observer frame in both $g$ and $r$ bands (see left panel of Fig.~\ref{fig:lc}), reaching an absolute peak magnitude of $-21.70$ in rest-frame $g$-band. A relatively close non-detections , and a low-order polynomial fit point toward to a rise time of $40\pm10$ days ($\sim$37d in rest-frame) from explosion. 
The rest-frame decline in $g$-band is 2.0 mag 100d$^{-1}$ (phase$<$10d), while in $V$-band is 4.5 mag 100d$^{-1}$ (phase$>$20d) suggesting a linear decline similar to that of SN2008es showing 2.9 mag 100d$^{-1}$ in $g$-band. This would imply SLSNe II fall on the right/top part of the $s2/M_{\rm max}$\footnote{We used the s2 parameter since we do not observe a transition from cooling to recombination phase, usually measured with the s1 parameter.} distribution of type II SNe \citet{an14a} and broadly follow the trend ($s2_{\rm SLSNe II}\gtrsim$2.9 mag and $M_{\rm max}(SLSNe II) >-19$ mag). This would support an association between SLSNe II and normal SNe II. The untimely end of our observational campaign in March 2013 was due to the right ascension of the object \footnote{Part of these data, mainly an early version of the rest-frame $g$-band and bolometric light curve, were previously given to \citet{ni15} as a courtesy.}. The second season started when it reappeared again in July and 
PESSTO restarted its annual observing periods in August 2013. Hence the follow-up campaign
recovered \hx\/ some five months after maximum light in this second season. 
From 160d after peak the light curves show a steady decline in all bands. 

\br\/ was detected during the rise phase, only in the $w$-band of PSST and by the Catalina Sky Survey (CSS). We continued to observe the SN until it disappeared behind the Sun in July 2015 and started again in December 2015.
We fit the data with a low-order polynomial and find the peak epoch consistent with MJD $57089.25\pm2.00$. Then, the non-detection of the transient $30$d before the first detection suggests a rise time of $\sim$35d in the rest-frame and hence similar to that of \hx\/ (see right panel of Fig.~\ref{fig:lc}  where $w$-band measurements have been converted in $r$-band through {\sc snap}, see Appendix~\ref{sec:s3}). However, the rest-frame peak magnitudes of $-20.22$ ($g$) and $<-20.75$ ($U$) are somewhat fainter than those of \hx\/ and \es\/ while its spectroscopic evolution is similar to them (see also Sections~\ref{sec:sp}~\&~\ref{sec:int}).
\br\/ shows a fairly rapid and linear decline in the UV and $u$ bands after peak, which is 
gradually less steep in the redder bands. In $B$, $V$, $g$ and $r$ bands, the light curves show a slower second decline with 1.2 mag 100d$^{-1}$ in $g$-band between 30-90 days post peak in rest-frame. Last two photospheric epochs of \br\/ could suggest a faster decrease similar to what experienced by some type II at similar phase \citep[see][]{va15}, which could correspond to the end of hydrogen recombination.   
In the second season (phase~$>250$d) the light curves show a very slow, almost flat, behaviour in all bands. This is the consequence of interaction between the SN ejecta and CSM material similar to that experienced by \hx\/ (see Section~\ref{sec:int}).

Due to \br\/ fainter peak luminosity with respect to the other two objects, it is important to understand how it compares with the few hydrogen-rich SNe populating a similar absolute luminosity space. \br\/ shows an overall slower evolution than other objects at similar absolute magnitude \citep[$\gtrsim -19.5$ in $g$-band,][]{ar16}, suggesting as \br\/ is different from the transients showing peak magnitude between those of normal type II and SLSNe. The only other non-SLSN object displaying a somewhat similar brightness is the type IIn/IIL SN2013fc \citep[M$_g \sim -20.2$,][]{kan16}, which however decreases of $>3.1$ mag 100d$^{-1}$ in $g$-band in the first 90 days. Moreover, SN2013fc displays a similar spectroscopic behaviour to type II suggesting a CSM interaction as source for its bright peak luminosity. Although \br\/ peak luminosity is closer to the range between SLSN and normal type II SNe, it shows a photometric behaviour and spectroscopic evolution different than the objects presented in \citet{ar16} and \citep{kan16}.

\begin{figure*}
\includegraphics[width=18cm]{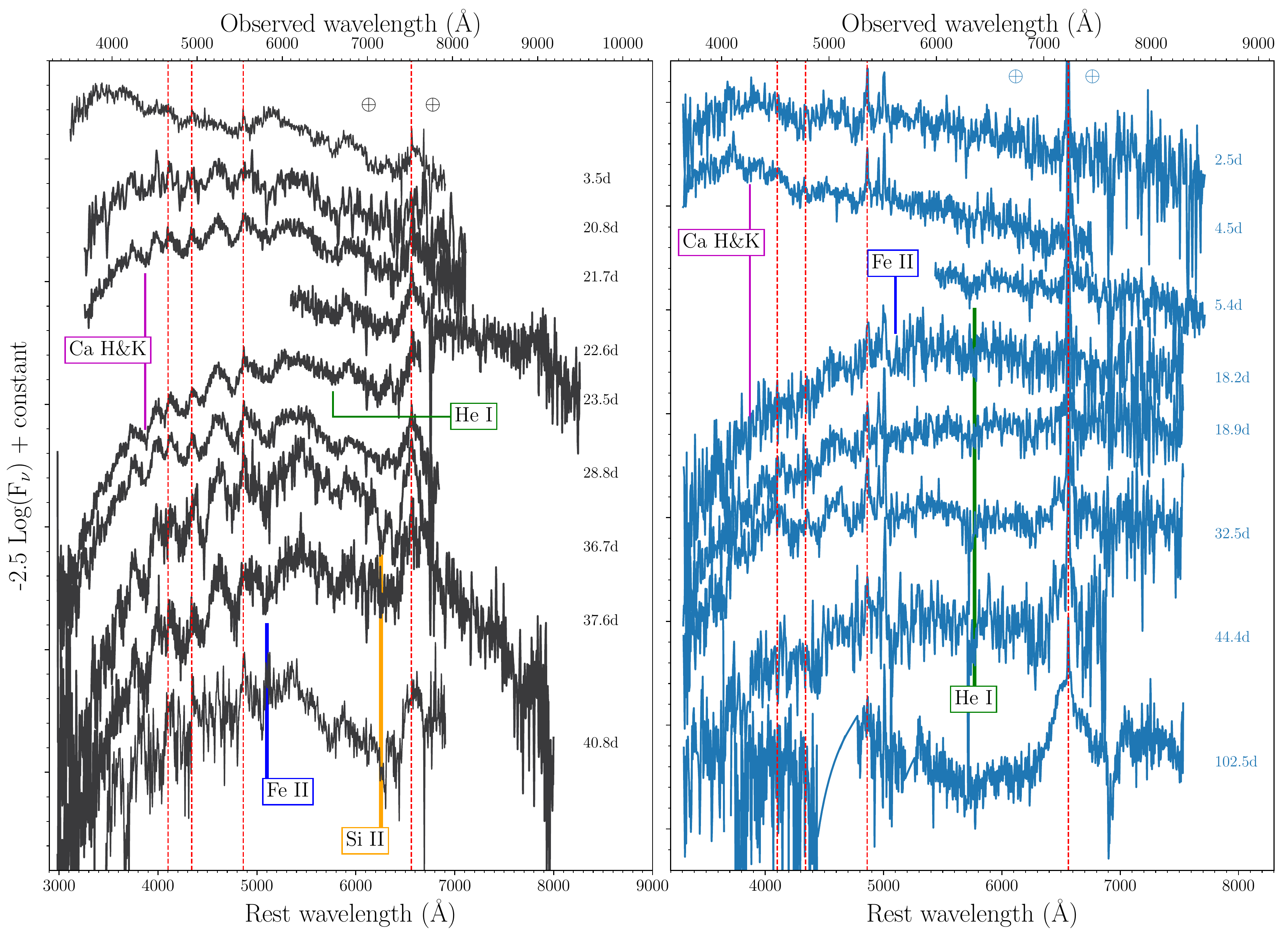}
\caption{Left: selection of photospheric spectra evolution of \hx\/. Right: selection of spectra evolution of \br. The phase of each spectrum relative to light curve peak in the rest frame is shown on the right. The spectra are corrected for Galactic extinction and reported in the rest frames. The most prominent features are labeled. Balmer lines \Ha\/ to H$\delta$ are marked with red vertical dashed lines. The $\oplus$ symbols mark the positions of the strongest telluric absorptions. WiFeS spectra were convolved with a factor of five and subsequently binned to a 5~\AA\/ scale. 
}
\label{fig:spev}
\end{figure*}

\subsection{Spectroscopy}\label{sec:sp}

The spectra evolution of the two SNe is shown in Fig.~\ref{fig:spev} (cfr. Table~\ref{table:sp}). Both \hx\/ (left panel of Fig.~\ref{fig:spev}) and \br\/ (right panel of Fig.~\ref{fig:spev}), like \es\/ \citep{ge09,mi09}, do not show the typical broad absorption features of O~{\sc ii}  
that we see in SLSNe I. We note that 
the spectra have temperatures consistent with the photospheric temperatures  
at which these ionic transitions are prominent 
\citep[$12000-16000$ K,][]{qu13,in13}, even though such ions could be sensitive to non-thermal radiation in a way similar to He~{\sc i} \citep{maz16}. 
In general the post peak spectra of both objects show Balmer lines, He, Ca, Fe and other metal lines.

\begin{figure}
\includegraphics[width=\columnwidth]{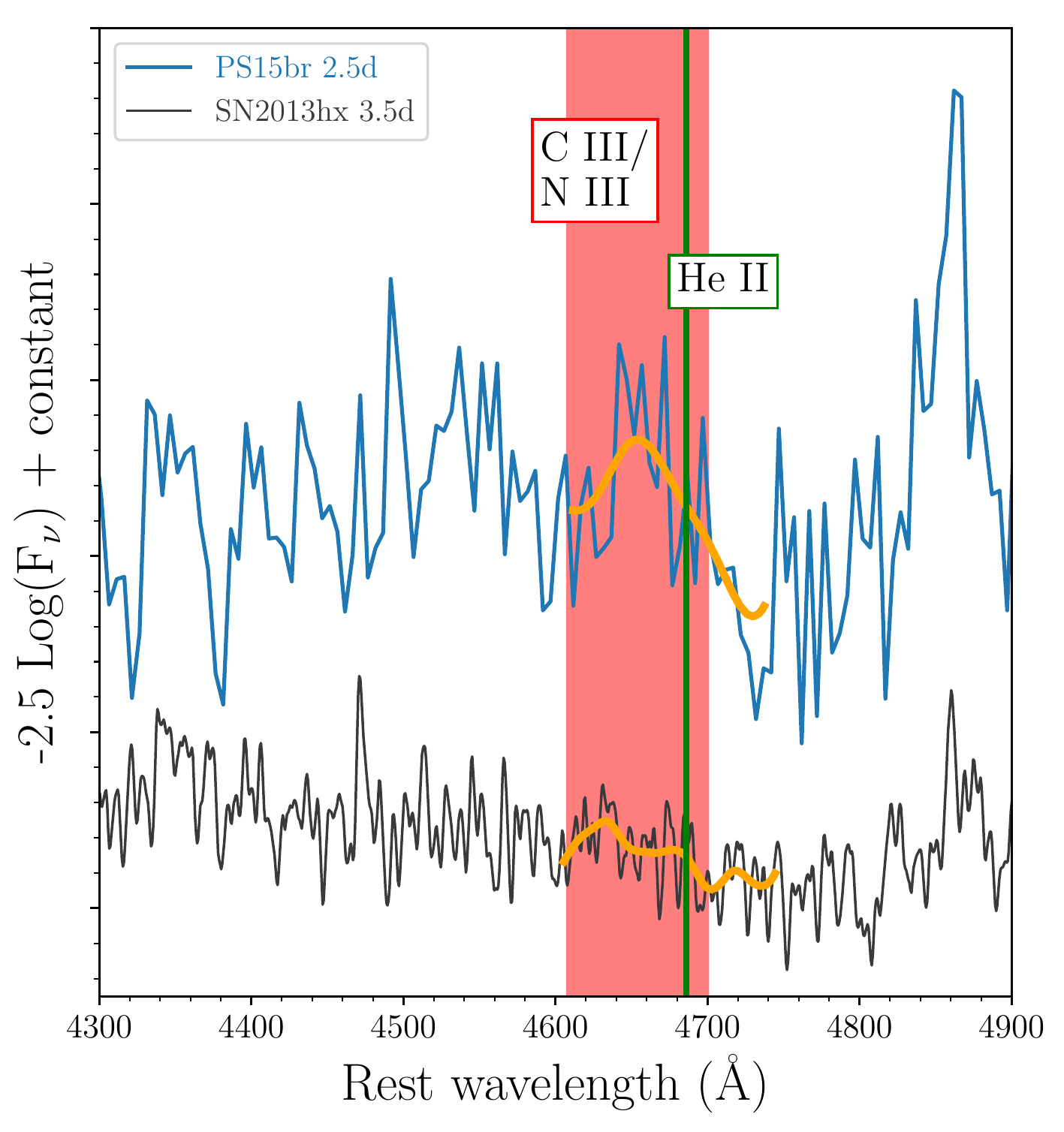}
\caption{First spectrum of each SN in the region around 4600~\AA\/ with the rest-frame He~{\sc ii} $\lambda$4686 marked by a green vertical line and the region of C~{\sc iii}/N~{\sc iii} lines by a red area. A gaussian filter (in orange) has been applied to highlight the wavelength region of interest. 
}
\label{fig:helium}
\end{figure}

The first spectrum of \hx\/, 3 days after peak, shows a broad \Ha\ and He~{\sc i} $\lambda$5876, while that of 21 days past peak also shows conspicuous Balmer lines and an emission feature at $\sim$4600~\AA\/  similar to those observed in \br\/ and \es\/ (see Fig.~\ref{fig:helium}). This feature was also present in the first spectrum but weaker. The emission peak of this feature could be again associated with C~{\sc iii}/N~{\sc iii}, possibly even contaminated by Fe~{\sc iii} forest lines rather than He~{\sc ii} $\lambda$4686 (see Fig.~\ref{fig:helium}). All the Balmer lines have a P-Cygni profile with the exception of \Ha\/ showing only the emission component up to $\sim30$ days after peak. However, the absorption component exhibited by \Ha\/ in the +36 to +38 day spectra is weak. Such behaviour resembles that of fast-declining type II SNe (or type IIL). A transient unresolved narrow \Ha\/ line tentatively detected in our first spectrum may point towards some weak interaction at early times, however such behaviour is seen in many type II SNe at early times \citep[e.g.][]{fa01,gy07,in13b,claudia17a}. A shallow He~{\sc i} $\lambda$5876 line, possibly blended with Na~{\sc iD}, is visible until 40 days after maximum.
The Ca H\&K lines are visible from +20.8 days, as well as a feature in absorption at $\sim$5000 \AA\/ likely related to the Fe~{\sc ii} multiplet $\lambda\lambda$4924, 5018, 5169.  Two other P-Cygni profiles are visible from 20 to 40 days after peak at 3745~\AA\/ and 4575~\AA\/ which, based on the similarities with type II SNe, we identify as Fe~{\sc i}. They are 
possibly blended with Ti~{\sc ii}, and Fe~{\sc ii}, respectively. 

An absorption feature is visible from +23d to +38d in \hx\/ (and less clearly from +5d to +19d in \br) on the blue side of \Ha\/ with an absorption minimum at 6250~\AA\/. This absorption feature has been observed in several type II \citep{chu07,in11,in12a,in12b,in13b,va14,va15,bo15} at both early or late photospheric epoch and goes by the name of `cachito' \citep{claudia17a}.
The two interpretations are high velocity hydrogen originating from interaction between rapidly expanding SN ejecta and a CSM - or  Si~{\sc ii} $\lambda$6355. The latter seems the most plausible identification in our objects since it shows a velocity consistent with, although slightly lower than, other metal lines such as Fe~{\sc ii} (see Section~\ref{ss:vel}). 

In the earliest spectrum of  \br\/  there is a feature at 4600~\AA\/ that could be associated with C~{\sc iii}/N~{\sc iii} rather than He~{\sc ii} $\lambda$4686 (see Fig.~\ref{fig:helium}). This feature already weakens in the spectrum at 5 days after maximum and has been observed in several non-superluminous type II \citep[e.g. SNe 1998S, 2007pk, 2009bw, 2013cu;][]{fa01,in12a,in13b,gy14}. \br\/ never shows a broad \Ha\/ absorption component but displays a multicomponent emission profile due to the combination of a relatively broad feature (v$\sim$1500 \kms) and a narrow component (v$\sim$110 \kms). This is likely due to the host galaxy since it has a similar width to those of the [O~{\sc iii}] emission lines of the host galaxy at $\lambda$4959 (v$\sim$110 \kms) and $\lambda$5007 (v$\sim$100 \kms) as measured from our WiFeS spectrum at early time (+4d) and Xshooter spectra at late time (337 and 362 days after maximum). This highlights that there is no evolution in the strength of the line and it supports their identification as galaxy line since are comparable to those of the host galaxies of SLSNe in previous studies \citep[e.g.][]{le15}.
\Hb\/ has a similar behaviour although the absorption component is visible from 20 days after maximum. 
Ca~{\sc ii} H\&K is visible in the first spectrum and becomes more prominent 
from day 18 when the metal lines appear as shown by the presence of the Fe~{\sc ii} multiplet $\lambda\lambda$4924, 5018, 5169 and possibly Fe~{\sc i} that replaces the 4600~\AA\/ feature. He~{\sc i} $\lambda$5876, or a combination of this element with Na~{\sc iD} $\lambda\lambda$5890, 5896, is detectable in the first spectrum and clearly from the third, showing a P-Cygni profile.

The last three \br\/ spectra of the photospheric phase (32 to 102 days) show the appearance of a broad component (FWHM~$\sim$8000 \kms) at \Ha\/ 
which is strongly asymmetric in the final photospheric spectrum at 102 days (see Fig.~\ref{fig:spev}).  The centroid of the broad \Ha\/ feature is somewhat obscured by the narrow galactic emission line, but is certainly blue shifted and we estimate its peak is at $-750$ \kms. 
The asymmetric blue emission is common for \Ha\/ of type II SNe and could be due to an opaque core concealing the outermost layers of the receding part of the envelope where the \Ha\/ is formed \citep{chu85}.
Alternatively it may be due to a steep density profile of the hydrogen layer, which increases the probability to observe blue-shifted photons since the opacity is mainly due to electron scattering and the line emission mostly comes from the region below the continuum photosphere \citep{de05,an14}.

Considering the slower light curve decline of \br\/ compared to \hx\/ and \es\/, the formation of the early \br\/ \Ha\/ profile is likely to be related to a weak interaction occurring at early epochs, whereas the SN ejecta would be responsible for the profile of late epochs. \br\/ is not a textbook interacting supernova since other ions, such as Fe~{\sc ii} and Ca H\&K, are observed with a P-Cygni profile during its photospheric evolution. The spectrophotometric early behaviour suggests an interaction with a nearby dense circumstellar medium that is swept up in the first week after maximum light. This behaviour is observed in many SNe II \citep{yaron17,morozova17,moriya17}, which show a transition from spectra with signs of weak interactions at early time to typical type II later on (14-30 days after explosion). Combining such spectroscopic evidences with the slow light curve decline, we suggest that \br\/ is a more luminous version of a transitional type II, but still different with respect to luminous IIn \citep[e.g.][]{smi07} showing a distinct spectrphotometric evolution.

The spectral evolution of these objects is similar (in the observed ions and their strength) and  provides additional information about hydrogen-rich SLSNe.

\subsection{Imaging polarimetry}\label{sec:pol}

On 2015 March 14 UT 05.2, corresponding to a rest-frame phase of 6 days after maximum, we obtained broadband polarimetry of PS15br with VLT/FORS2 using the $V$ filter \footnote{Unfortunately only one epoch was obtained due to a combination of decreasing apparent magnitude and ending of the ESO semester.}.
We observed it in four different half-wave retarder plate angles: 0, 22.5, 45 and 67.5 deg. 
The exposure time was 150~s for each angle.
The data were reduced in a standard manner by using bias and flat frames, without polarisation units in the light path.
We obtained PSF photometry of PS15br and six field stars in the field of view (FOV) and determined their Stokes parameters. 
The measurements were corrected for instrumental polarisation \citep{pr06} and interstellar polarisation (ISP).
For more details on the reductions and analysis, see \citet{le15b}.

\begin{figure}
\includegraphics[width=\columnwidth]{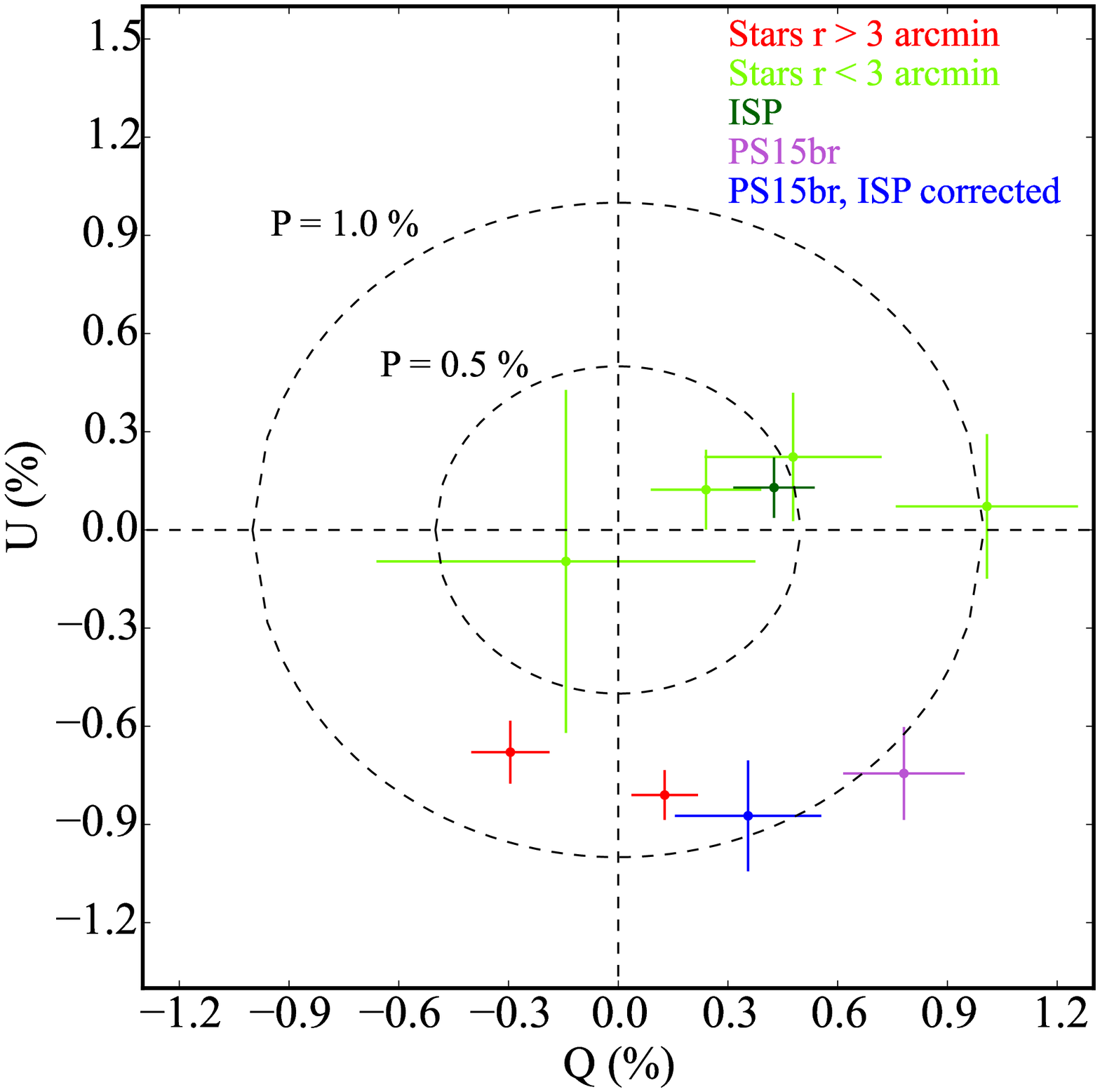}
\caption{ Q-U plane of PS15br at +6d from maximum light. Concentric circles show polarization degrees of 0.5\% and 1.0\%.}
\label{fig:pol}
\end{figure}

Fig.~\ref{fig:pol}, shows the location of PS15br and the field stars on the $Q-U$ plane, after correcting for instrumental polarisation. 
The error bars that are shown are the measurement errors resulting from the PSF fitting and essentially depend on the S/N ratio ($\sim$500 for PS15br).
The field stars (light green and red points) present a significant scatter in this plane and the SN signal (purple) is comparable to the one of the field stars. 
It is therefore likely that the signal of the field stars and the SN is dominated by ISP.
\citet{pr06} showed that the instrumental-induced polarisation in FORS1 - an instrument with identical design to that of FORS2 - has a radial pattern with polarisation increasing as we move away from the optical axis. 
The two field stars that are the main cause of the scatter (red) are at the outskirts of the FOV, more than 3.6\arcmin\/ from PS15br, which is in the centre of the FOV. 
On the other hand, the measurement of the stars closer to the object (light-green) are self-consistent with each other and can be used for a more accurate determination of the ISP.

The ISP is shown in Fig~\ref{fig:pol} as the weighted average of the light-green stars. 
Therefore, despite the fact that all points have been corrected for instrumental polarisation - using the FORS1 relation in \citet{pr06} - it is possible to obtain a self-consistent picture by only using field stars closer to the optical axis and the SN. These result in a consistent determination of the ISP ($Q_{\rm ISP} = 0.42 \pm 0.11$\%, $U_{\rm ISP} = 0.13 \pm 0.09$\%), and are clearly different from the properties of PS15br ($Q_{\rm SN} = 0.78 \pm 0.17$\% and $U_{\rm SN} = -0.74 \pm 0.14$\%.

By correcting for the ISP so evaluated, we obtain a significant polarisation signal $Q = 0.36 \pm 0.20$\% and $U = -0.87 \pm 0.17$\%, resulting in $P = 0.94 \pm 0.17$\%. After correcting for polarisation bias, according to \citet{pr06}, this value reduces to $P_0 = 0.93 \pm 0.17$\%.
Indicatively, this level of polarisation would correspond to an asymmetry of E~$\sim$~10-15\%, assuming an ellipsoidal photosphere \citep{ho91}. Such level of polarisation is consistent with that reported for SLSN I SN2015bn at similar epochs \citep{in16a,le17} but higher than that of SLSN I LSQ14mo \citep{le15b}, for which no evidence for significant deviation from spherical symmetry was found. On the other hand, such measurement is higher than those reported for type II SNe soon after peak \citep[see][for a review]{ww08} and a factor of two less than that of type IIn \citep[e.g. SN1998S,][]{wang01}, which are however dominated by strong H emission lines not visible in \br.

However, of we assume that all field stars are equally good for the determination of the ISP, we conclude that there is a significant scatter - probably pointing to an unaccounted source of systematic error - and that no conclusive evidence on the source of the polarisation associated with PS15br can be derived. Unfortunately, there are no other suitable stars (bright and not saturated) in the FOV of PS15br to resolve this uncertainty.

\section{SLSNe II into context}

\begin{figure}
\includegraphics[width=\columnwidth]{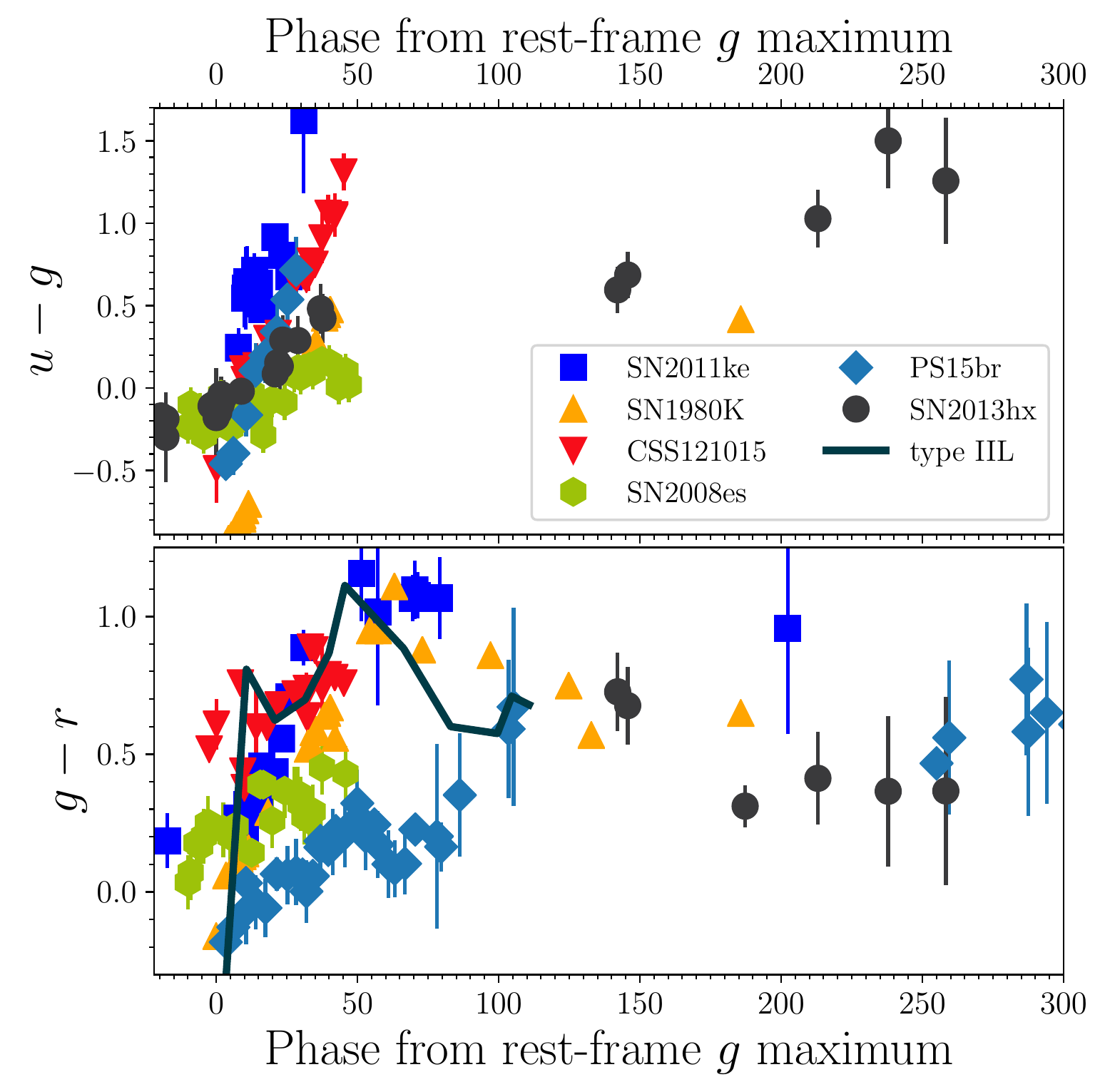}
\caption{Comparison of the dereddened and $K$-corrected colour evolutions. \hx\/ and \br\/ are shown together with the other SLSN II SN2008es \citep{ge09,mi09}, SLSNe~I SN20011ke \citep{in13} and CSS121015 \citep{ben14} together with type II SN1980K, for which the colour curve is actually $U-B$ (top panel) and $B-V$ (bottom panel), and that of a type IIL template from \citet{sanders15}.}
\label{fig:ccomp}
\end{figure}

\subsection{Colour curves}\label{sec:cc}
Since colour curves are useful probes of the temperature evolution of
supernovae, we computed the rest-frame curves, after accounting for the
reddening and redshift effects of time-dilation and $K$-correction.
In Fig.~\ref{fig:ccomp}, \hx\/ and \es\/ show a constant $u-g$ increase toward redder values.  
However, \br\/ is $\sim0.25$ mag bluer than other SLSNe II at peak. Its colour evolution 
is also steep, but similar to the prototypical SLSN I SN2011ke and CSS121015, a SLSN I showing early interaction with a H-shell. This could support the suggestion of an early, weak interaction in \br\ as observed in several transitional type II.
We also note that after 20 days, the $u-g$ colour evolution gradient of \hx\/ is similar to that of fast-declining type II (or type IIL, $u-g\sim0.4$ mag), whereas \br\/ is slightly steeper and \es\/ slightly shallower.
Although the former is, overall, shifted redwards. 

SLSNe II (\br\/ and \es) $g$-$r$ colours show a similar behavior with a slow, monotonic increase toward redder colours until 50 days. SLSNe II do not show the constant colour exhibited by SLSNe I from the pre-peak phase to $\sim$15d \citep{in13}. In general the colours evolve more slowly to the red than type IIL SNe, which cool down to $g-r\sim$0.6 in the first 20 days as shown by a template colour evolution \citep{sanders15}. We note that although such template is derived from type II, peak luminosity and light curve evolution of the objects are similar to type IIL similarly to other sample papers using Bessell filters \citep[e.g.][]{li11,faran14,galbany16,valenti16}. In addition, it appears that \br\/ $g-r$ 
shows a turnover at 50 days and evolves bluewards before
turning to the red again. However, considering the uncertainties and compare the colour curve with the temperature behavior (see Section~\ref{ss:vel}) it may be a flattening in the colour evolution 
(at  $g-r\sim0.24$ mag) rather than a blueward evolution. 
The object then clearly evolves to the red afterwards and
with $g-r=0.6$ mag after 100d it appears similar to \hx\/ and type II SNe. After 200 days \hx\/ and \br\/ colour curves remain steady at $g-r=0.4$ mag and $g-r=0.6$ mag in a similar fashion to that experienced by normal type II/IIL such as SN1980K.

\subsection{Bolometric luminosity}\label{sec:bol}

UV to NIR photometry is required to
obtain a direct measurement of the full bolometric luminosity. This is
typically difficult to achieve at all epochs during a SN evolution. 
Despite the lack of data, valid corrections can be applied to the observed photometric bands to
compute the total bolometric flux \citep[e.g.][for a further insight]{pasto15,ch15,kan16,in17}.

\begin{figure}
\includegraphics[width=\columnwidth]{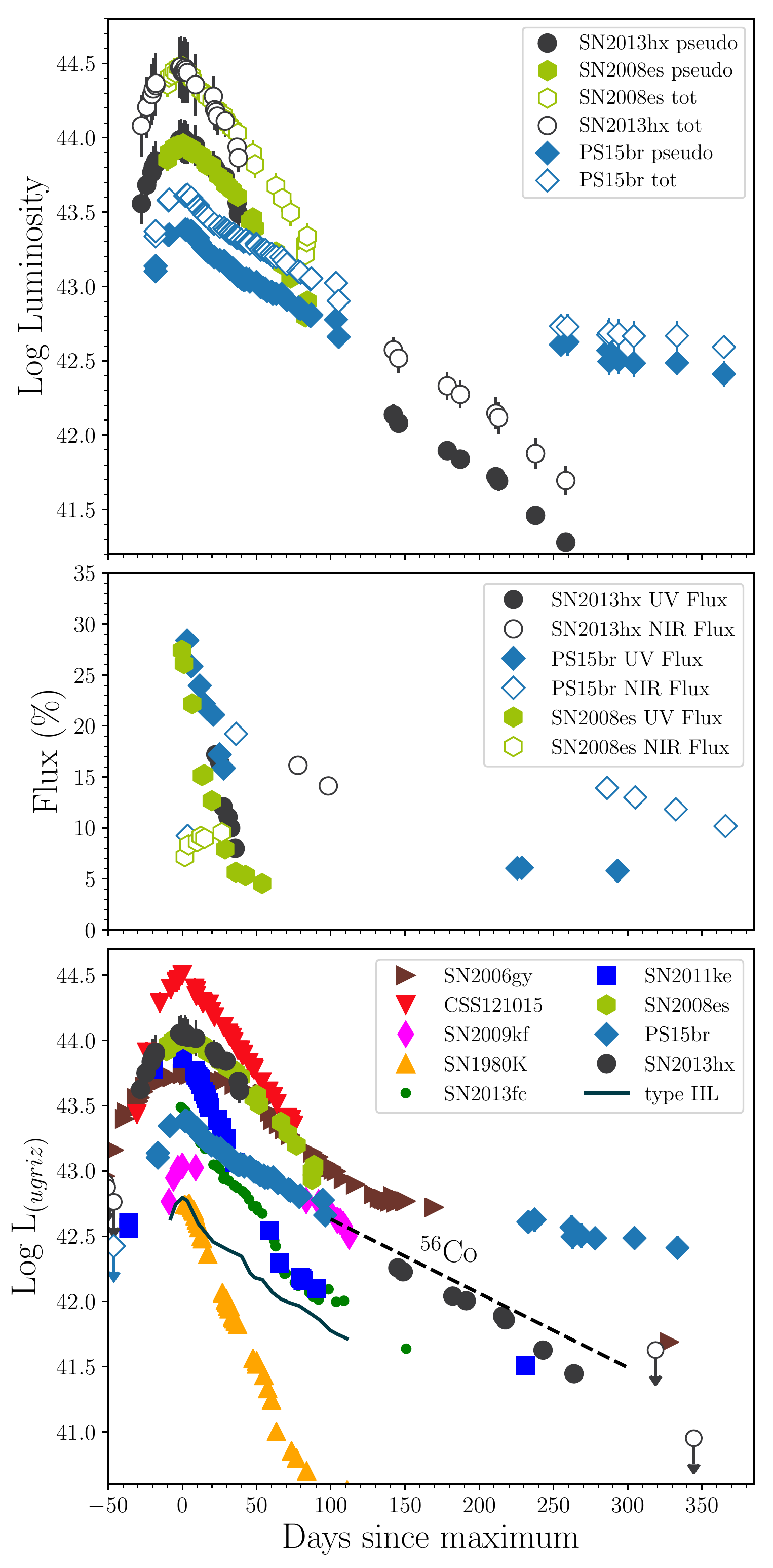}
\caption{Top: pseudo (filled symbols) and proper bolometric (open symbols) of three SLSNe II. Middle: percentage of the bolometric flux in UV (filled symbols) and NIR (open symbols) with respect to the total bolometric flux derived from data of SLSNe II. Bottom: comparison of pseudo bolometric light curves ($ugriz$) of \hx\/, \br\/ and SLSN II 2008es \citep{ge09,mi09}, the interacting SLSN IIn 2006gy \citep{smi07}, the prototypical SLSN I SN2011ke \citep{in13} and the early interacting SLSN I CSS121015 \citep{ben14}, the type II SNe 1980K \citep{ba82k} and 2009kf \citep{bot10}, the bright and mild interacting type II 2013fc \citep{kan16}, as well as a type IIL template \citep{sanders15}. These bolometric light curves are computed after correcting the observed broadband photometry for time dilation and applying $K$-corrections, through {\sc snake} (see Appendix~\ref{sec:s3}), and extrapolating up to $z$ band when this was unavailable. The black dashed line is the slope of \co\/ to \fe\/ decay.} 
\label{fig:bolom}
\end{figure}

The effective temperatures of the photosphere of SLSNe II during
their first 30-50 days after explosion are between $T_{\rm bb} \sim
10000 - 16000$\,K (see Section~\ref{ss:vel}). This means
that their fluxes peak in the UV ($\lambda<3000$~\AA) during this
period while our rest-frame $ugri$ bands typically cover from
3500~\AA\/ redwards.  Thus a significant fraction of the flux is not
covered by the optical $ugri$ imaging, as testified by the UV to total luminosity ratio for \es\/ and \br\/ (see middle panel of Fig.~\ref{fig:bolom}).  At around 20d after peak, the
effective temperatures tend to drop below 10000 K, hence the SEDs peak
between 3000\,\AA\ and 4000\,\AA. Although the peak of the SED moves
redward, a significant amount of the bolometric flux is radiated in
the UV even during these late stages. In the following, we will use the term
`pseudo-bolometric light curve' to refer to a bolometric light curve
determined using only the optical filters with
the flux set to zero outside the observed bands. 

To calculate bolometric luminosity, the broad band magnitudes in the available optical bands were converted into fluxes at the effective filter wavelengths, then were corrected for the adopted extinctions (cfr. Section~\ref{sec:sample}). A SED was then computed over the wavelengths covered and the flux under the SED was integrated assuming there was zero flux beyond the integration limits.  Fluxes were then converted to luminosities using the distances previously adopted. We initially determined the points on
the pseudo-bolometric light curves at epochs when $ugri$ were available simultaneously (or very close in time) and later for epochs with coverage in less than the four filters.  Magnitudes from the missing bands were
generally estimated by interpolating the light curves using low-order
polynomials ($n\leq3$) between the nearest points in time.  For some points this
interpolation was not possible and hence we extrapolated the magnitudes assuming constant colours from neighbouring epochs.

Since \hx\/ and \es\/ have similar optical light curves we built the \hx\/ bolometric light curve considering the same NIR flux contribution of \es\/ at early epochs (\es\/ broad band data were retrieved by \citet{mi09,ge09}). To check the validity of this assumption we also estimated the bolometric light curve integrating, from UV to NIR wavelengths (1000~\AA\/ -- 25000~\AA\/), our best blackbody fit to the available \hx\/ SED and found similar NIR flux contribution to the bolometric light curve to that of the previous method. The difference between these two methods is included as uncertainty in the evaluation of the \hx\/ bolometric light curve. In Fig.~\ref{fig:bolom} (middle panel) we compared UV and NIR contributions to the total bolometric luminosity of \hx\/ and \br\/ with that of \es. The UV contributions are similar and drop almost to zero after 60 days from maximum, while the NIR contribution increases with time. We have late NIR data in $JHK$ information for \hx\/ and \br, whereas  \es\/  only has data in the  $J$ and $H$
 bands at these epochs where the NIR contribution becomes
significant. If we were to ignore the  \br\/ $K$-band contribution, 
the two percentages of NIR contribution ($JH$ of \br\/ and \es) are similar.
Therefore when we evaluated the full bolometric light curves of \hx\/ and \es\/ we added the $K$-band contribution estimated specifically by \br\/ NIR data.

The peak luminosities of our bolometric light curves are 
$\mathrm {L_{\rm \hx}}\approx2.75\times10^{44}$ erg s$^{-1}$,
$\mathrm {L_{\rm \br}}\approx4.15\times10^{43}$ erg s$^{-1}$ and
$\mathrm {L_{\rm \es}}\approx3.00\times10^{44}$ erg s$^{-1}$. 
While the maximum luminosities reached by the pseudo-bolometric light curves are almost factor two lower.
We note that \es\/ values are in agreement with those previously reported by \citet{ge09,mi09}.

The bottom panel of Fig.~\ref{fig:bolom} compares the pseudo-bolometric light curves $ugriz$ of our sample of SLSNe II - namely \hx, \es, \br\/ - with that of the peculiar SLSN~I CSS121015,
a prototypical SLSN I (SN2011ke) and type II SNe. We note that the extrapolation up to $z$ band has been done assuming constant colour from neighbouring epochs or integrating the best fit blackbody to the available SED up to the $z$ band coverage. The comparison remarks that light curves of  \hx\/ and \es\/ are notably similar. CSS121015 also exhibits a fairly similar decline over the first 40 days, but is 0.5 dex brighter than any other SN. \hx\/ rise appears slower than typical SLSN I events like SN2011ke. \br\/ is $\sim0.5$ dex fainter than the other two SLSNe II and sits in the gap between normal luminosity type II SNe and SLSNe II. Although \br\/ exhibits a similar decline to \es\/ and \hx\/ over the first 30 days, it clearly does not show the faster decline observed after this epoch - it remains on quite a linear decline for nearly 100 days. During this time baseline it also shows a similar overall decline rate similar to that of one of the brightest type II, SN2009kf \citep{bot10}. \br\/ is slower than the interacting SN2013fc \citep{kan16}, which is the only type II reaching a luminosity similar to those of \br\/ due to its early interaction, \br\/ decline of $\sim0.4$ dex over the first 40 days after peak is similar to that shown by the type IIL template on a similar timescale, but slower afterwards. 

\begin{figure}
\includegraphics[width=\columnwidth]{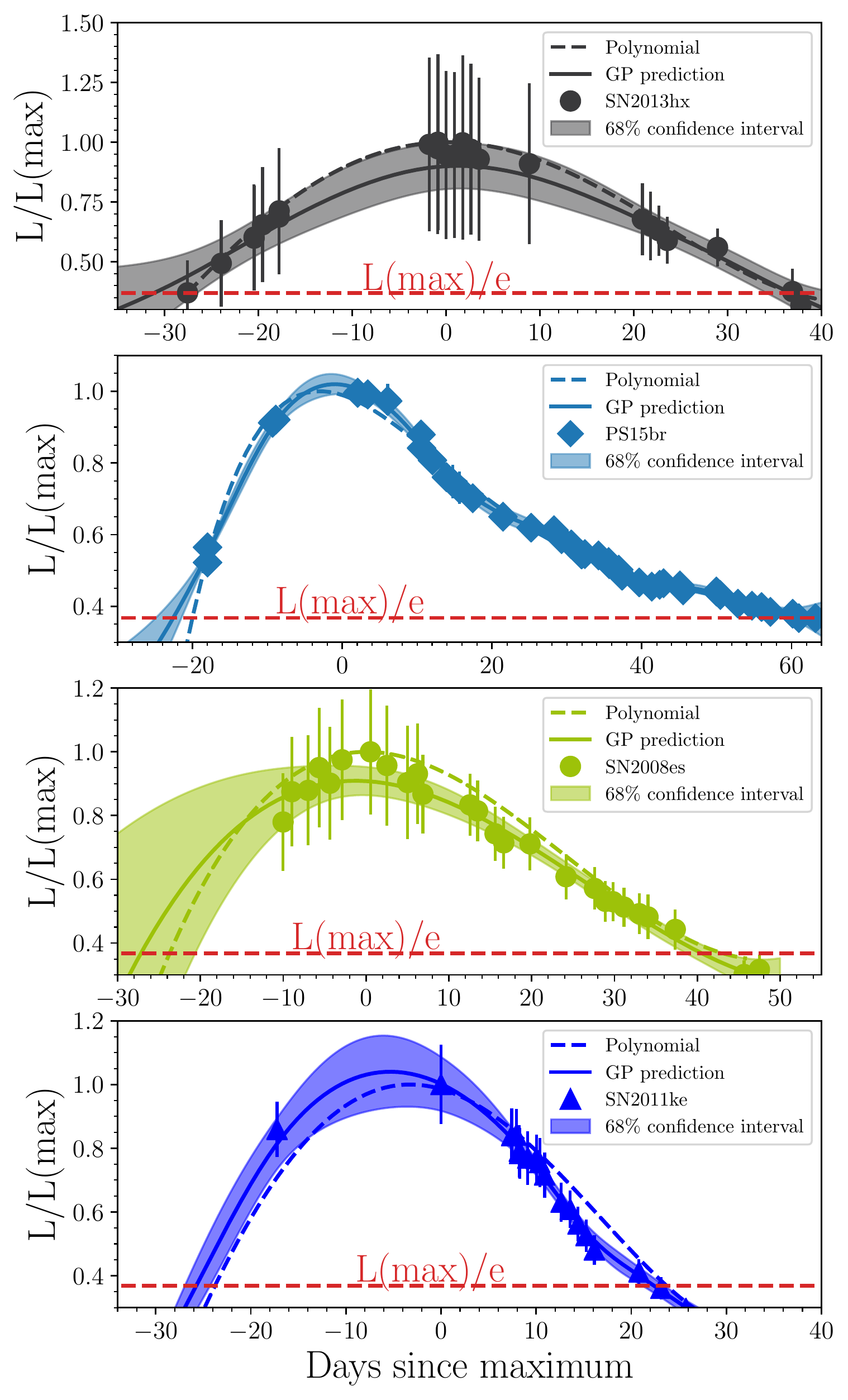}
\caption{Pseudo bolometric light curves ($ugriz$) normalized to the maximum of the data, which is at zero days. The luminosity equivalent to L(max)/e is highlighted by a red dashed line. We note that the GP peak for SN2008es and SN2011ke occurs earlier in time than that of the polynomial and data}
\label{fig:bw}
\end{figure}

\begin{table*}
\caption{Width of SLSNe pseudo-bolomotric $ugriz$ light curves. P refers to the polynomial measurements, while GP to those performed with Gaussian processes.}
\begin{center}
\begin{tabular}{lccc|ccc}
\hline
\hline
SN & width (P) & rise (P) & decline (P) & width (GP) & rise (GP) & decline (GP) \\
& (day) & (day) & (day) & (day) & (day) & (day)\\
\hline
\hx\/ & $64\pm3$& 27 & 37& $69\pm8$& $31\pm7$ & $38\pm4$\\
\br\/ & $80\pm5$& 20& 60& $83\pm4$& $23\pm4$ & $60\pm4$\\
\es\/ & $67\pm5$&24 &43& $70\pm16$& $30\pm16$ & $40\pm4$\\
SN2011ke& $48\pm3$&24 &24& $47\pm4$& $18\pm3$ & $29\pm2$\\
\hline
\label{table:bolp}
\end{tabular}
\end{center}
\end{table*}

The comparison shows that both 
\hx\/ and \es\/ have 
fairly similar shaped profiles within 10-20 days of peak. 
To make quantitative comparisons of the
light curve widths, we measured
a width when the pseudo-bolometric flux is $1/e$ less than that at maximum light in order to match the estimates of the diffusion timescale parameter $\tau_{m}$ in \citet{ni15}.
When these epochs were not specifically covered by a data point, we initially used low order polynomial fits ($n\leq4$) to interpolate and find the width (see Fig.~\ref{fig:bw}). The values 
are reported in Table~\ref{table:bolp}. We note that the values are slightly different than those reported in \citet{ni15}, that could be due to small difference in the evaluation of the bolometric light curves. The uncertainties reported are estimated from the differences between the values achieved with different polynomials orders and time baselines. We also employed Gaussian processes (GP), which are generic supervised learning methods widely used in the machine learning community \citep{bishop,gps} for Bayesian regression and classification problems and successfully used in the context of SN \citep[e.g.][]{kim13,scalzo14,2017arXiv170901513D} and SLSN light curve fitting \citep{in17c}. An important advantage of GP regression compared to other regression techniques is that it produces a best-fit model together with uncertainty at each point and a full covariance estimate of the result at unknown points \citep{sdm2014}. We used the machine learning packages {\sc george} \citep{george} written in {\sc python} and a Matern 3/2 kernel to fit our light curve width \citep[see][for further details about the methodology, the kernel and package choice]{in17c}. The results are reported in Table~\ref{table:bolp}, while a comparison with the polynomial fitting can be seen in Fig.~\ref{fig:bw}. Since the two methods seems overall comparable, hereafter we will refer to the GP results and not those of the polynomial fitting.

As highlighted before, the decline of SLSNe II tend to be slower than that of SLSNe I such as SN2011ke (a prototypical SLSN I), while the rise time is comparable for all objects. The $1/e$ width is $\gtrsim70$ days for SLSNe II and around 50 days for SN2011ke. We used the same techniques to measure the width of CSS121015 (SLSN I interacting with a hydrogen shell) and found it to be similar to that of SN2011ke. That is a consequence of the sharp rise and decline from peak luminosity observed in the bottom panel of Fig.~\ref{fig:bolom}. We also note that \hx\/ and \es\/ have a similar decline from 20 to 50 days.  
The width of the main peak in \br\/ is also not comparable to that of the SLSN IIn SN2006gy \citep{smi07}, as the latter has a much wider light curve of $\sim$110 days  (see bottom panel of Fig.~\ref{fig:bolom}). Furthermore, the \br\/ peak is strongly asymmetric (see Fig.~\ref{fig:bolom}, Fig.~\ref{fig:bw} \& Table~\ref{table:bolp}), a detail conveying against a strong interaction with a hydrogen-rich massive shell typical of IIn (and SLSN IIn) which would have resulted in a wide, almost symmetric light curve peak. 
 The slower decline experienced by \br\/ is comparable to that of the bright type II SN2009kf and could be the consequence of a more massive hydrogen layer  - and hence higher contribution from hydrogen recombination - than those of the other SLSNe II. Alternatively it could be due to interaction in a similar manner to what has been observed in weakly or mildly interacting type II such as SNe 2007pk \citep{pritchard12,in13b}, PTF11iqb \citep{smi15} and 2013fc \citep{kan16}. These SNe can be explained with an interaction, weaker than of SNe IIn, with a dense CSM most likely caused by wind acceleration \citep{morozova17,moriya17}, which affects spectra and light curve evolution in a less dramatic way than what observed in SN IIn/SLSN IIn such as SN2006gy.

\subsection{Spectroscopic resemblance to other SLSNe and type II SNe}

\begin{figure}
\includegraphics[width=\columnwidth]{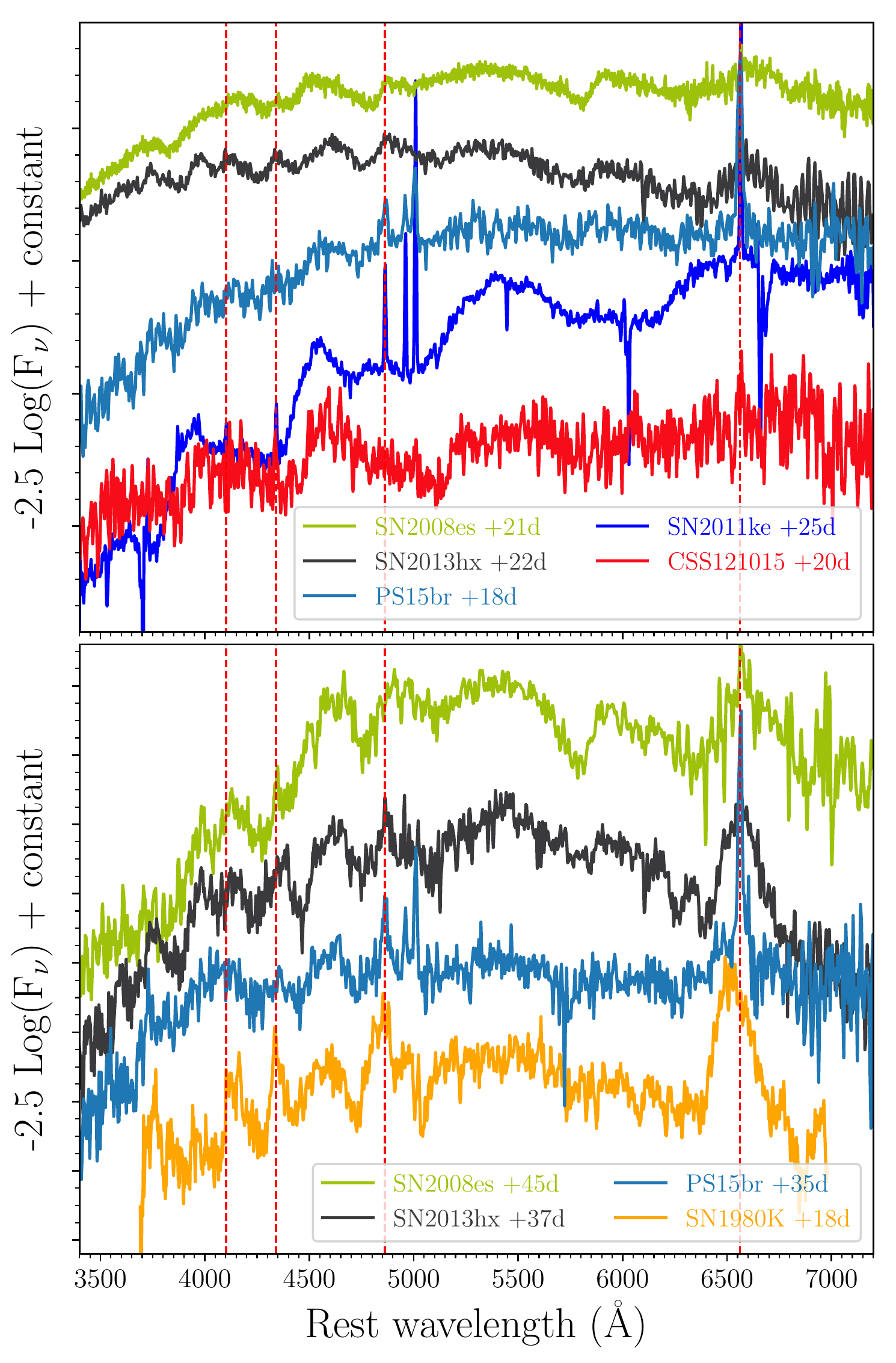}
\caption{Top: comparison in Log(F$_{\nu}$) space, which highlights spectral features, of \hx\/ and \br\/ spectra at $\sim$20 days from maximum with those of SLSNe showing Hydrogen in their spectra SN2008es~\citep[][]{mi09} and CSS121015 \citep{ben14}, together with SLSN I SN2011ke \citep{in13}. Bottom: Comparison of the spectra of \hx\/ and \br\/ at $\sim$40 days from maximum together with that of SN2008es at similar epoch and that of the type II SN1980K at $\sim$20 day after maximum light. The spectra of the three SLSNe II are similar to that of the type II in a similar fashion to SLSNe I at $\sim$30d resembling type Ic at maximum. \Ha\/ to \Hd\/ are marked with red vertical dashed lines in all panels.}
\label{fig:spcmp}
\end{figure}

\begin{figure*}
\includegraphics[width=18cm]{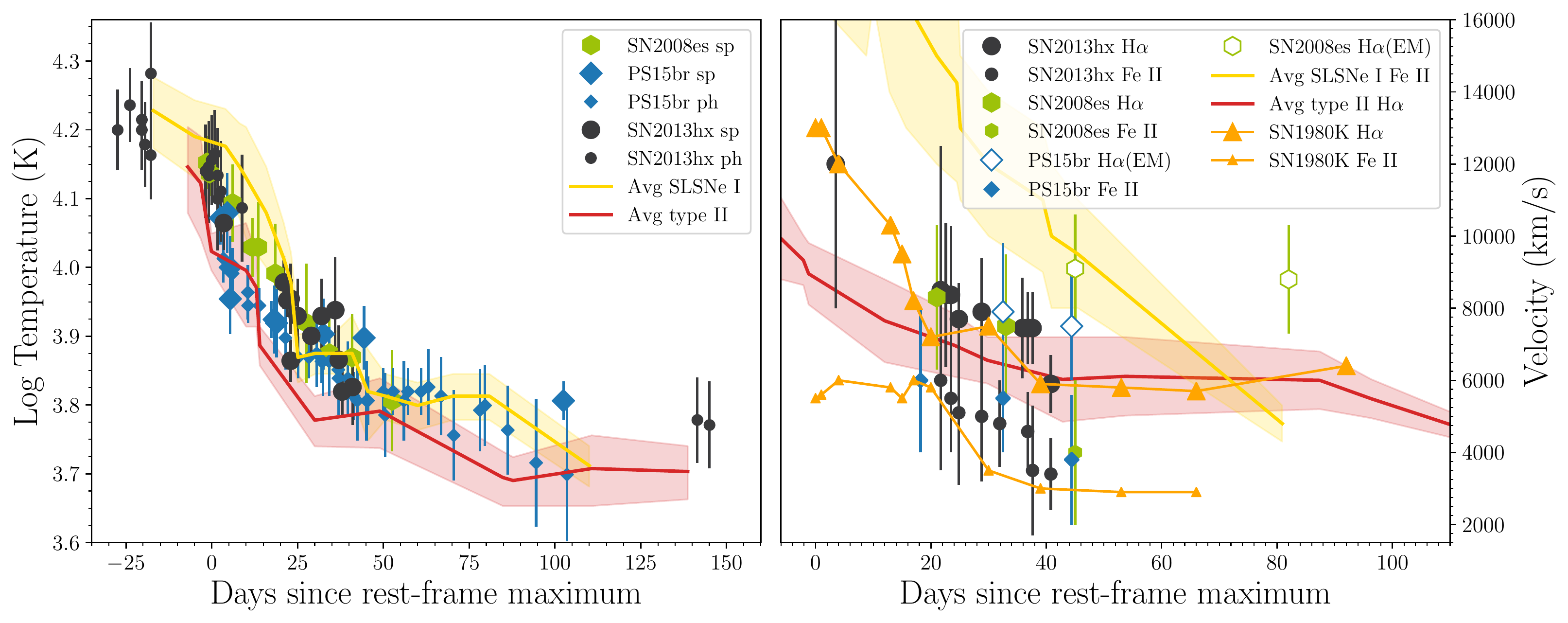}
\caption{Left: \hx\/  (black) and \br\/  (blue) temperature evolution compared with that of SN2008es (green). The gold region represent the 99.73\% region of the temperature space of SLSNe I \citep{in13,ni15}, while the red that of a type II collection \citep[][]{el03,ma10,in11,in12a,bot10,tom13,claudia17b}. Right: \Ha\/ and Fe~{\sc ii} evolution of \hx\/ (black) and \br\/ (blue) compared with that of SN2008es (green), the type II SN1980K \citep[orange triangle and line for \Ha\/ and smaller symbols for Fe~{\sc ii};][]{ba82k}. The gold region represents the velocity space of Fe~{\sc ii} line in SLSNe I, while the red that of \Ha\/ in type II SNe. Open symbols refer to velocities computed with the FWHM of the emission lines.}
\label{fig:velTevol}
\end{figure*}

In the top panel of Fig.~\ref{fig:spcmp} the comparison of the spectra around +20d post maximum with those of the other SLSNe such as SN2008es, SN2011ke and CSS121015 highlights a difference in the line evolution between SLSNe II and both a prototypical (SN2011ke) and peculiar (CSS121015) SLSN I. CSS121015 shows weaker H than SN2013hx and SN2008es with an equivalent width (EW) EW(\Ha)$_{\rm SN2013hx,SN2008es}\sim 10 \times {\rm EW(H\alpha)}_{\rm CSS121015}$.
CSS121015 also exhibits Fe~{\sc ii} lines which are more prominent than the type II SLSNe at similar phase. Indeed the Fe absorption lines in CSS121015 are stronger than those of SLSNe II \hx\/, \br\/ and \es\/ with EW(Fe)$_{\rm CSS121015}\approx {\rm EW(Fe)}_{\rm SLSNe I} \sim 5-10 \times {\rm EW(Fe)}_{\rm SLSNe II}$. Both \hx\/ and \br\/ are also different from typical SLSNe I spectra. These show the same metal lines observed in CSS121015 at similar epochs.
The difference between SLSNe I, dominated by broad metal lines, and SLSNe II (\es, \hx, \br), exhibiting H lines with weaker metal lines with respect to the SLSNe I, is highlighted in the top panel of Fig.~\ref{fig:spcmp}.
To investigate if SLSNe II spectroscopically behave in a similar fashion of what observed for SLSNe I \citep{pasto10,in13} in Fig.~\ref{fig:spcmp} (bottom panel) we compare the spectra of \hx\/ and \br\/ with that of \es\/ at around 40 days, together with that of the type II SN 1980K\footnote{Some of such spectra were digitalised and flux calibrated in Benetti PhD thesis, 1992.} at 20 days past maximum light. There is an overall resemblance between SLSNe II and SN1980K. At these epochs SNe II/IIL exhibit $v\sim$8200 \kms\/ 
components in \Ha\/ 
while \hx\/  shows $v\sim$8000 \kms. Similar trends are visible for  \Hb, with $v$(\Hb)$_{\rm SN1980K}\sim7900$ \kms, which is similar to  $v$(\Hb)$_{\rm SN2013hx}\sim7000$ \kms. In addition,  the strength of the He~{\sc i} $\lambda$5876 line is similar between the objects (EW$_{\rm SN1980K}=10$\AA\/ compared to EW$_{\rm SN2013hx}=11$\AA\/, while EW$_{\rm SN2008es}=18$\AA\/ ).  This also holds true for the Fe~{\sc ii} feature, which has  EW$_{\rm SN1980K}=16$\AA\/ and EW$_{\rm SN2013hx}=18$\AA\/. Blueward of \Hb\/ the profiles begin to differ, mainly due to Fe~{\sc i}/Ti~{\sc ii} shown around 4000~\AA\/ in SLSNe and a stronger Ca H\&K  (EW$_{\rm SLSNeII} \sim 2-3 \times {\rm EW}_{\rm SN1980K}$)
The line profiles of \br\/ are similar to those of \hx\/ and \es\/, with the only exception of the \Ha\/ profile that does not show an absorption component and a rather weak He~{\sc i}. This support the theory that a certain degree of weak/mild interaction is at play in \br\/.
\br\/ does not show a blue pseudo-continuum, which would result in a rise of the flux bluewards than $\sim$5400~\AA\/ \citep[e.g.][]{tu00,ben14,in16b} that characterises certain types of IIn spectra or the blue spectra with only lorentian Balmer emission line typical of SLSNe IIn such as SN2006gy \citep{smi07}. Indeed, bluewards than \Ha\/, \br\/ is similar to \hx\/ and \es\/ in line strength. 
 
\subsection{Velocity and temperature evolution}\label{ss:vel}

In the left panel of Fig.~\ref{fig:velTevol} the evolution of the temperature is plotted in logaritmic space. It is derived from the blackbody fit to the continuum of our rest-frame spectra (big symbols) and colour temperatures (small symbols). 
Only \hx\/ has multi-band coverage before peak allowing us to estimate the temperature pre-peak (17000 to 15000~K, $\sim4.2$ dex). \hx\/ shows a monotonic decline from 14000~K (4.15 dex) at peak to 6600~K (3.82 dex) at 40 days, while \br\/ declines from 12000~K (4.08 dex) to 6000~K (3.78 dex) showing similar temperatures to \hx\/ and \es\/ in the same time coverage. 
After 50 days, the temperatures of the objects flatten at $\sim$5500~K (3.74 dex). \es\/ shows a similar temperature evolution to that of \hx\/. The overall temperature evolution of \hx, \br\/ and \es\/ is similar to those observed in SLSNe I and normal type II SNe with a similar decline to those of type II. We note that in the first 10 days after maximum \br\/ shows a drop in temperature similar to those of type II and \hx, but $\sim0.1$ dex fainter than the latter.

In the right panel of Fig.~\ref{fig:velTevol} the expansion velocities for our SLSNe, as inferred from the position of the absorption minima of the P-Cygni profiles for \Ha\/ and Fe~{\sc ii} (average of $\lambda\lambda$4924, 5018, 5169)  are compared with those of the only other SLSN II \es\/, an average of SLSNe I and type II SNe. The uncertainties were estimated from the scatter between several measurements. In the case of \Ha\/,  the velocities were computed from the FWHM of the emission lines for \br\/ and for SN2008es in spectra after 40 days.
\Ha\/ velocities decline monotonically from 12000 \kms\/ to 7500 \kms\/ on the time scale available for \hx. An almost identical decrease is observed in \es\/ and \br\/ in the available period (20 $<$ phase (day) $<$ 45) even if \br\/ values are due to the broad emission component.
The decline rate is qualitatively similar to normal luminosity type II SNe which have average velocities that are lower by about 1200 \kms\/ at similar epochs, whereas SN1980K \Ha\/ velocity evolution resembles those of SLSNe II. The velocities measured from the full width at half maximum (FWHM) of the emission components appear to be systematically higher than those inferred from the minima. The Fe~{\sc ii} velocities are usually assumed to be better tracers of the photospheric velocity and they 
are similar in SN2013hx, SN2008es and \br\/ decreasing from 6000~\kms\/ to 4000~\kms\/  \citep[we note that our measurements are in agreement with those presented by][]{ni15}. These show the same decline as displayed in normal faster-declining type II, but 
with a shift or delay by of $\sim$15-20 days. SLSNe II Fe~{\sc ii} velocities are quite different from those of SLSNe I, in that they are $\sim$7000 \kms\/ slower. 

\begin{figure*}
\includegraphics[width=18cm]{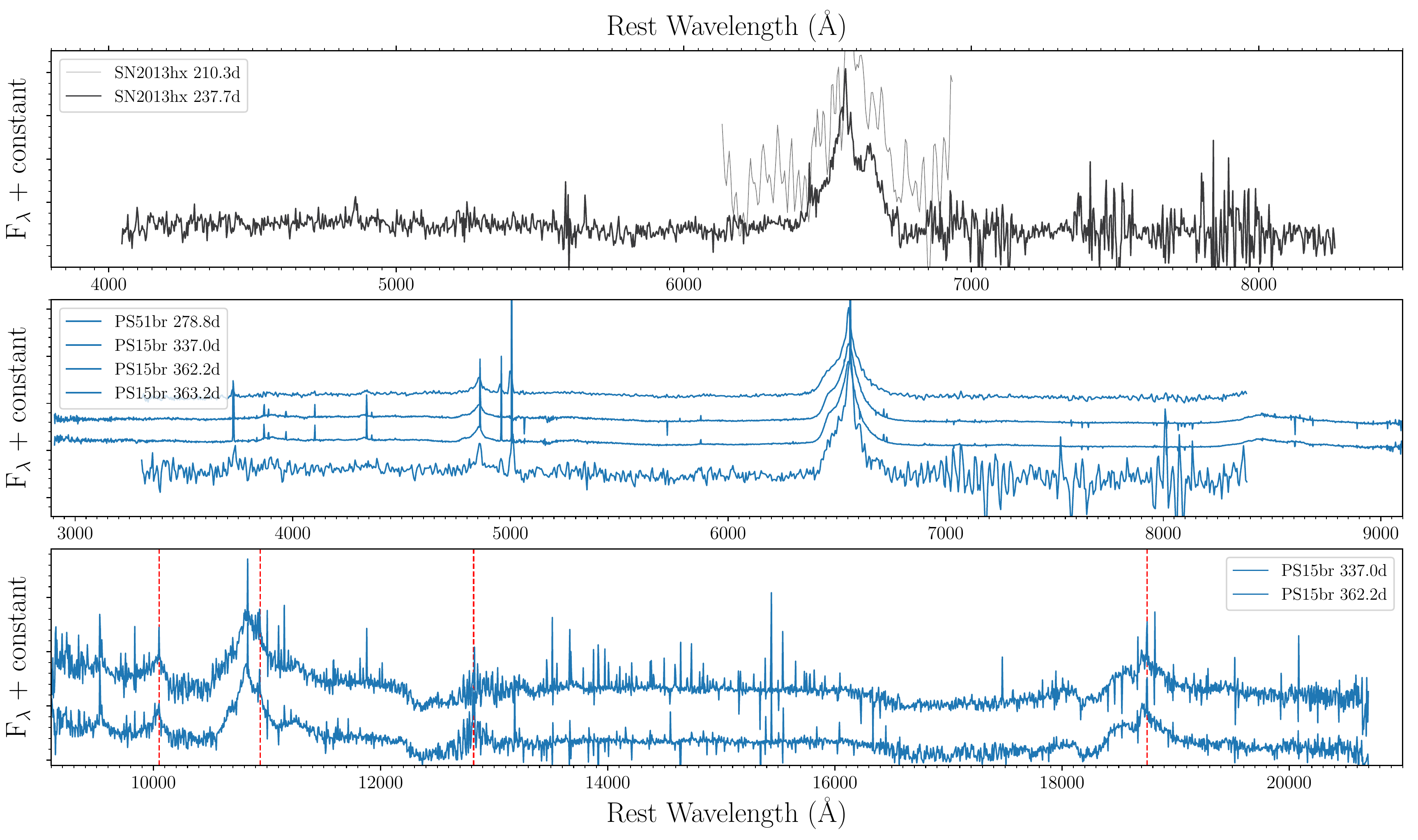}
\caption{Top: Last two spectra of \hx\/. Middle: Late time optical spectra of \br\/. Bottom: Late time NIR spectra of \br\/. Paschen lines from $\alpha$ to $\delta$ are identified by vertical dashed red lines.}
\label{fig:last}
\end{figure*}

\begin{figure*}
\includegraphics[width=18cm]{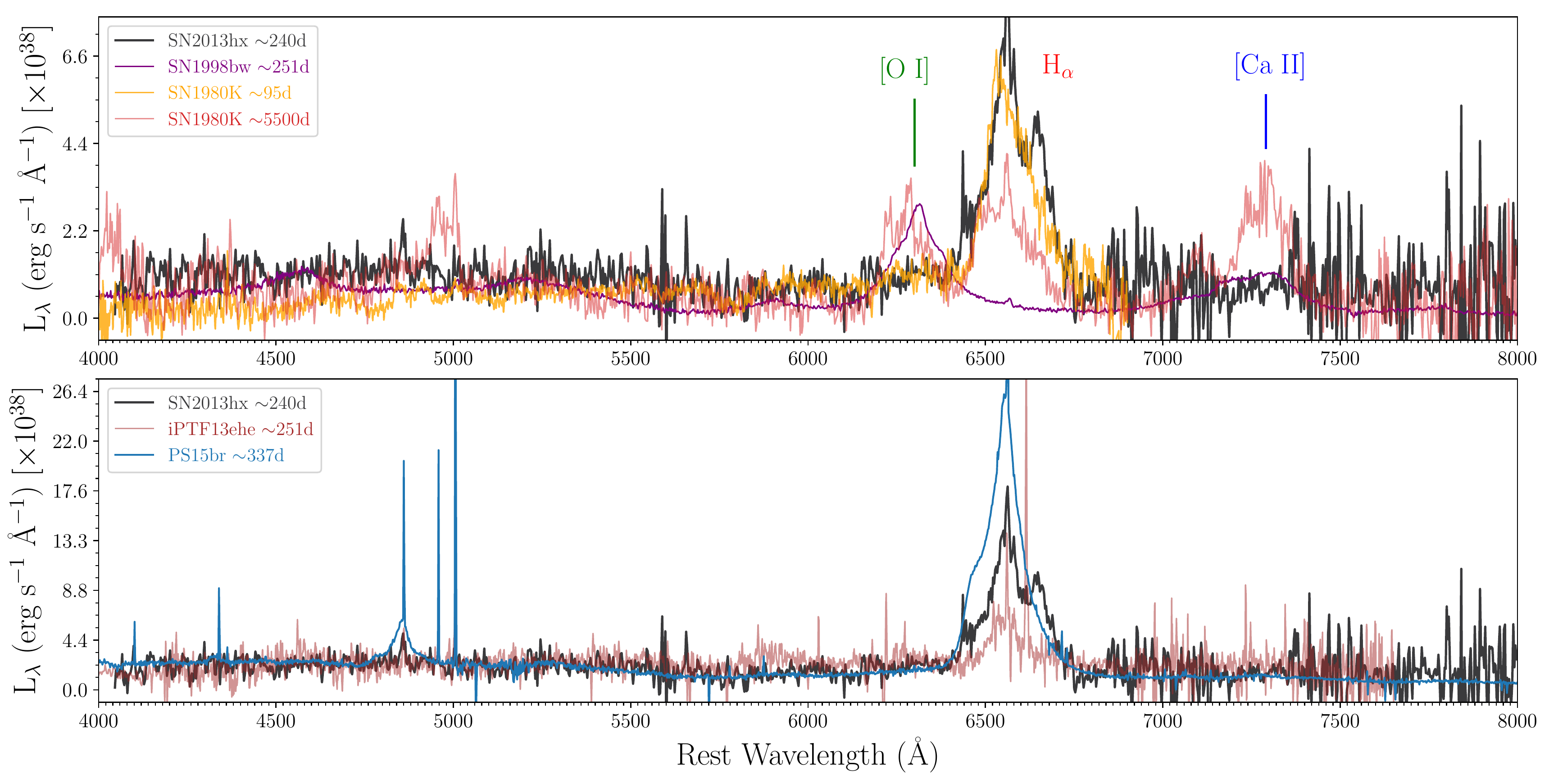}
\caption{Top: the VLT+FORS2 spectrum of \hx\/ compared with late spectra of the type IIL SN1980K \citep{ba82k} and a late spectrum of the broad-line type Ic SN1998bw \citep{pat01}. All the spectra have been scaled to the continuum flux level of \hx. The most prominent line of nebular core-collapse spectra are labelled. Bottom: the VLT+FORS2 spectrum of \hx\/ and the VLT+XShooter (337d) compared with SLSNe I iPTF13ehe, a slow SLSN I showing late \Ha\/ emission.}
\label{fig:lastcmp}
\end{figure*}

\section{Late time spectra and signs of interaction in SLSNe II}\label{sec:int}

In Fig.~\ref{fig:last} we show the late VLT+FORS2 spectrum of \hx\/ and those of \br\/ obtained with VLT+Xshooter and NTT+EFOSC2, which are the latest spectra and the only at late ($>$150d from peak) phase for SLSNe II obtained to date.   
For \hx\/, a lower signal-to-noise spectrum was taken
by PESSTO on +210d, and is also shown in Fig.~\ref{fig:last}. This NTT spectrum does not appear to be 
significantly different to the FORS2 +237d spectrum and as it is much lower quality, we will not discuss it further. In the case of \br\/, PESSTO spectra do not show any substantial evolution with respect to those of XShooter and have lower S/N, hence we will focus our analysis only on the XShooter spectra.

There is no distinguishable line at the position of  [O~{\sc i}] $\lambda\lambda$6300, 6364\AA, which is usually prominent in core-collapse supernovae. There is perhaps some hint of weak and broad emission in \hx\/ that could be attributed to this line, but the the detection is $\lesssim2\sigma$ with respect to the continuum flux and is not convincing (see Fig.~\ref{fig:lastcmp}). 
In Figure~\ref{fig:lastcmp} we compared our \hx\/ spectrum with those of the fast-declining type II (or type IIL) SN1980K and the broad-line type Ic SN1998bw at late epochs and scaled to the same luminosity distance of \hx\/ (570.3 Mpc) in order to match the rest-frame spectral luminosity exhibited by SN2013hx.
The spectra of SN2013hx and SN1980K are similar, although they differ by 135 days with respect to the peak epochs, and as consequence also similar to that of PS15br shown in the bottom plot. 
The overall similarity with type IIL SNe suggest that also SLSNe II are consistent with a star explosion similar in mass to those of normal core-collapse SNe. 
The \Ha\/ profile and the blueward features are also different from that of iPTF13ehe, a SLSN I with late-time interaction with a hydrogen shell. The width of the central and blue components of SN2013hx \Ha\/ are comparable with those of PS15br at $\sim337$d.
PS15br H$_{\beta}$ emission component is stronger than those almost non-existent of SN2013hx and SN1980K.

\begin{figure}
\includegraphics[width=\columnwidth]{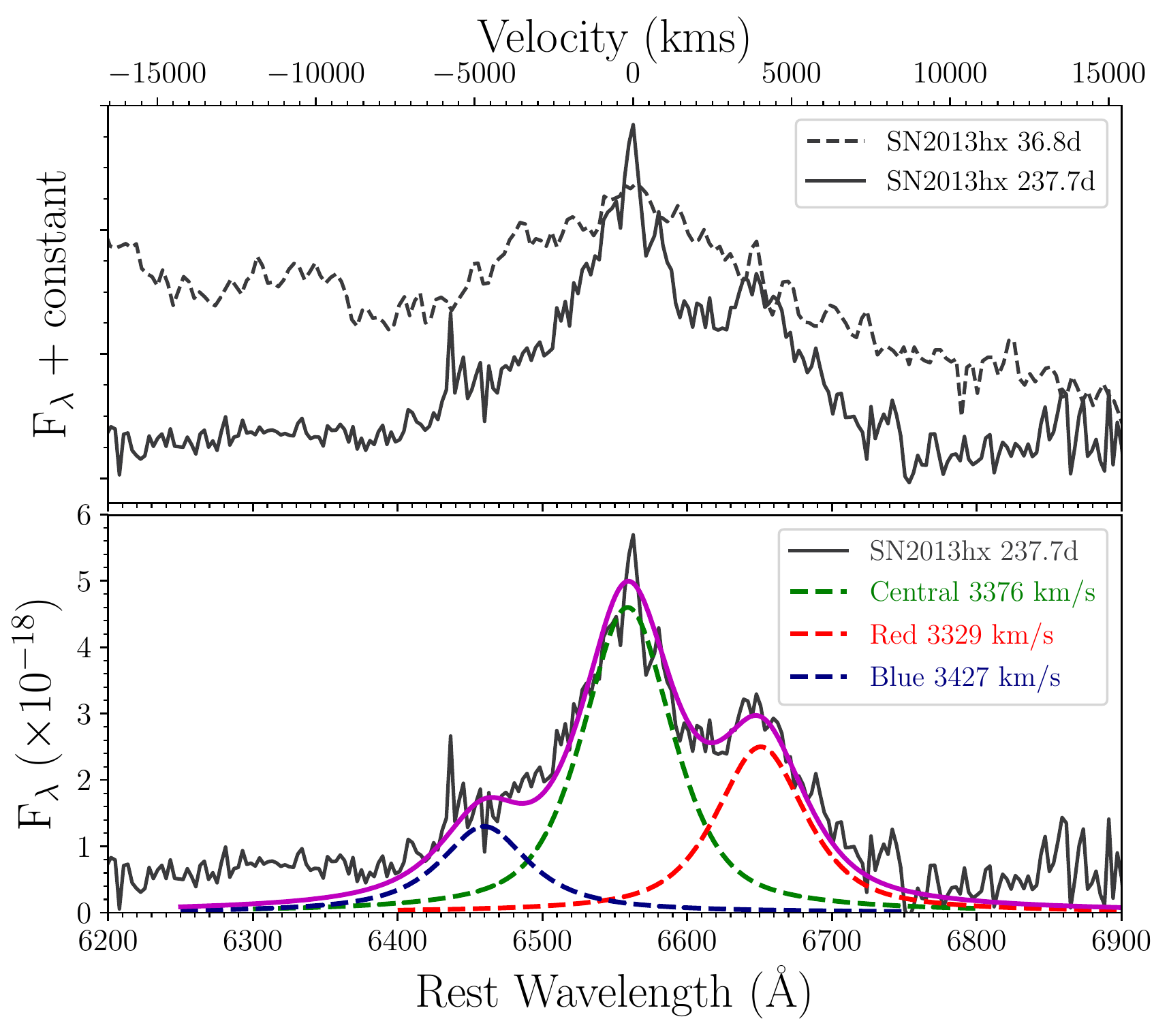}
\caption{Top: \Ha\/ multicomponent profile of \hx\/ spectrum at 238 days (black solid line) with that exhibited at 37 days (black dashed line) scaled in flux to match the previous spectrum. Bottom: \Ha\/ multicomponent profile of the spectrum at 238 days since maximum. The spectrum is dominated by interaction with a H-rich CSM.}
\label{fig:fit}
\end{figure}

\begin{figure}
\includegraphics[width=\columnwidth]{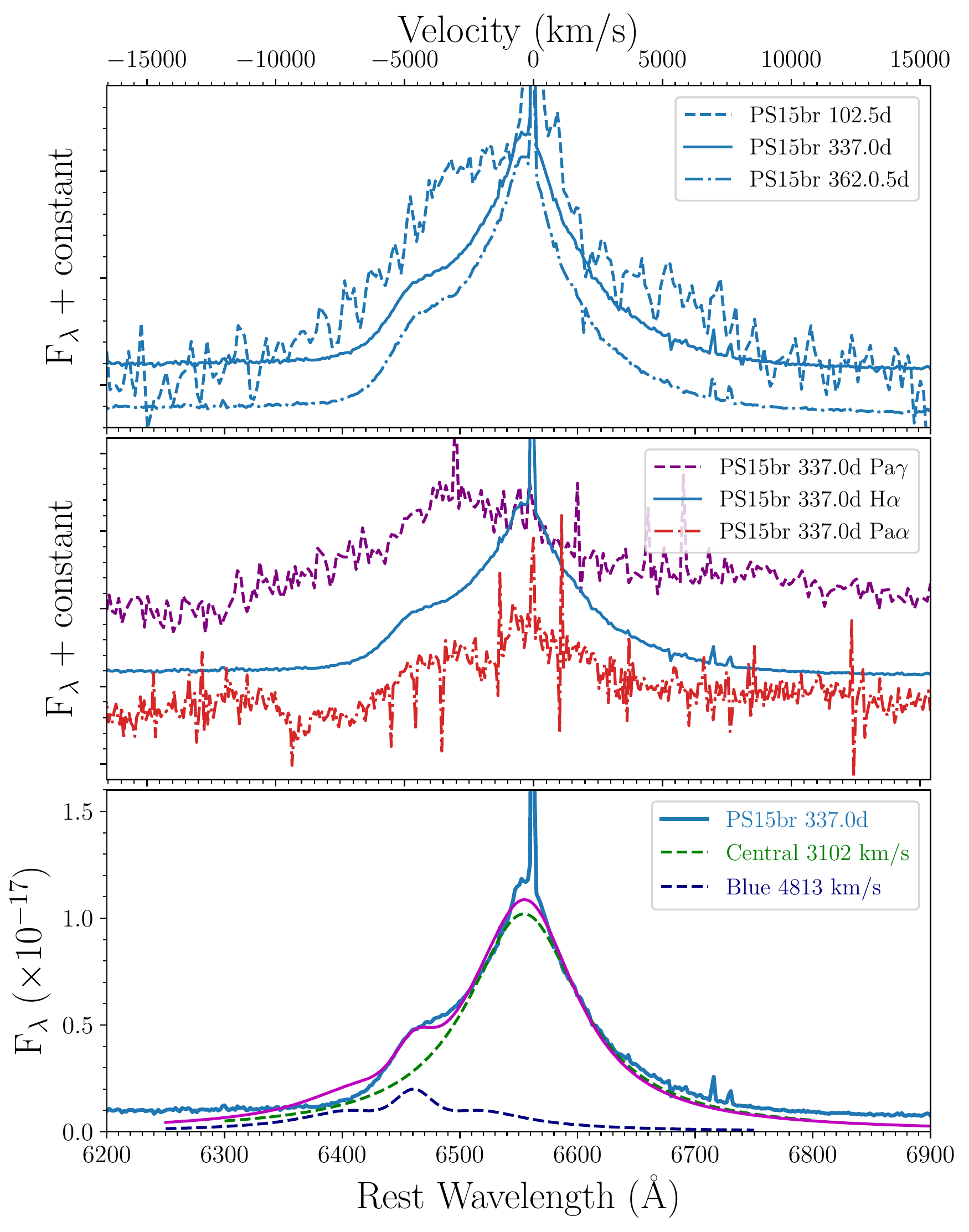}
\caption{Top: \Ha\/ multicomponent profile of \br\/ spectrum at 337 and 362 days (blue lines) with that exhibited at 102 days (cyan line) scaled in flux to match the previous spectrum. Middle: \Ha\/ (blue line), Pa$\gamma$ (magenta line) and Pa$\alpha$ (red line) multicomponent profiles of the \br\/ spectrum at 337 days. Bottom: \Ha\/ multicomponent profile of the \br\/ spectrum at 337 days since maximum. The spectrum is dominated by interaction with a H-rich CSM.}
\label{fig:fitbr}
\end{figure}

The +247d  \hx\/ spectrum and all the late \br\/ spectra are dominated by a strong and multicomponent \Ha\/ profile, and the simplest explanation  would be that the SN ejecta is interacting with an H-rich, highly asymmetric CSM \citep[as previously suggested for similar late time interactions; e.g.][]{ben16}.  There are no other lines visible in the optical with the exception of \Hb\/, and the 
continuum is faint. However, as shown in Fig.~\ref{fig:last} (bottom panel), \br\/ NIR spectra exhibit Paschen lines in emission, as well as He~{\sc i} $\lambda$10830 that obscures Pa$\gamma$. We note that both \Hb\/ and Pa$\alpha$ profiles resemble that of \Ha\/ with a prominent blue shoulder.
In \hx\/ the \Hb\/ flux is $\sim$50 times less than that of the \Ha\/ profile and $\sim$30 times less than \Ha\/ central component (see Table~\ref{table:intera}). For case B recombination in the temperature regime $2500 \leq {\rm T (K)}\leq 10000 $ and electron density $10^2 \leq n_e \leq 10^6$ the \Ha\/ line should be 3 times stronger than \Hb\/. However, the case B recombination is never observed in type II before a couple of years since both the $n=4\rightarrow n=2$ (\Hb) and $n=3\rightarrow n=2$ (\Ha) transitions are optically thick \citep{1992ApJ...386..181X}. In this case the P$\alpha$ ($n=4 \rightarrow n=3$) transition will depopulate the $n=4$ state, with the photons escaping to produce
the P$\alpha$ emission line. \Ha\/ however is produced strongly as the only other alternative for 
depopulation of the the $n=3$ state 
is Lyman $\beta$. 
In \br\/ the \Hb\/ flux is $\sim$10 times less than that of the \Ha\/ profile and as expected the P$\alpha$ emission line is visible.
The integrated luminosity of the overall profile of \Ha\/ line 
are $\sim$9 and $\sim$20 times those of a normal type II at similar phase \citep{kf98} for \hx\/ and \br, respectively. This suggests an additional contribution to \Ha\/ luminosity in both cases.
What powers the decline rate of the late time light curve is an open question. One
could speculate that the signature of interaction we see at this late phase is indicative of the 
same powering source for the light curves between 100$-$260\,days. 
Late interaction has been observed in bright type IIL such as SNe 1979C, 1980K and 1986E \citep[][and references therein]{fe99,cap95,ben16}.

The \hx\/ \Ha\/ profile in Fig.~\ref{fig:fit} has a triple peak structure spanning 7400 \kms\/ across the base (cfr. Tab~\ref{table:intera}), which is similar to the triple peak shown by the interacting type II PTF11iqb after 500 days \citep{smi15}.  The central component is consistent with having zero rest-frame velocity, and
the blue component peaks at $-4700$ \kms\/ with respect to this \Ha\/ rest-frame velocity. This blue
component appears to be the weakest  of the three. The red component is at $+4000$ \kms. We analysed the feature with gaussian, lorentzian and pseudo voigt line profiles having set {\it a priori} the number of components (three) but letting vary the width and intensity of each component.  We retrieved the best fit with three voigt profiles with widths of $\sim$3400 \kms. However, gaussians have a similar goodness of fit  and produce similar line widths but somewhat different line fluxes. We performed the same analysis on the 337 days spectrum of \br. In this case, \Ha\/ shows a two components profile, a blue at $-4700$ \kms\/ w.r.t. \Ha\/ and the central at almost zero rest-frame velocity ($-390$ \kms). This profile is similar to that shown by \hx\/, with the exclusion of the red component, and it is also shown in the Pa$\alpha$ (see middle panel of Fig.~\ref{fig:fitbr}) but not in Pa$\gamma$, which is blended with the more prominent He~{\sc i} $\lambda$10830. The best fit for these components is retrieved with  voigt profiles of width $\sim$3100 and $\sim$4800 \kms\/ for the blue and central component, respectively (see Fig.~\ref{fig:fitbr}).

\begin{table*}
\caption{Late \Ha\/ profile measurements of SLSNe II \hx\/ ($\sim$240 day) and \br\/ ($\sim$337 day)}
\begin{center}
\begin{tabular}{lccccc}
\hline
\hline
Feature & Peak & Width & Width & Velocity w.r.t. \Ha\/ peak & Flux\\
& (\AA) & (\AA) & (\kms) & (\kms) & (erg s$^{-1}$ cm$^{-2}$)\\
\hline
\multicolumn{6}{c} \hx \\
\hline
Blue & $6459.8$& 73.8 & 3427& $-4700$ &1.1$\times10^{-16}$\\
Central & $6558.9$& 73.8& 3376 &$-190$ &4.0$\times10^{-16}$\\
Red  & $6650.8$&73.8 &3329 & $4000$ &2.2$\times10^{-16}$\\
\hline
\multicolumn{6}{c} \br \\
\hline
Blue & $6460.7$& 66.8 & 3102& $-4700$ &2.4$\times10^{-16}$ \\
Central & $6556.5$& 105.2& 4813 &$-320$ & 9.3$\times10^{-16}$\\
\hline
\label{table:intera}
\end{tabular}
\end{center}
\end{table*}

In general, the line profile could be schematically explained by:
\begin{enumerate}
\item a relatively broad emission from the SN ejecta with FWHM $\sim$3400 \kms\/; 
\item interaction of the H-rich ejecta with:
\begin{itemize}
\item[(a)] a ring/disk edge-on,
\item[(b)] a dense, clumpy CSM;
\end{itemize}
\item a likely unshocked CSM that could explain the narrow, unresolved emission on the top of the central component.
\end{enumerate}

The blue wing of the \Ha\/ emission of \hx\/ and \br\/ roughly coincides in wavelength with the corresponding blue wing of the \Ha\/ profile at 37 and 102 days as shown by the top panel of Figs.~\ref{fig:fit}~\&~\ref{fig:fitbr}. If the absorption minimum maintains the same velocity shown at early time, it will be at a similar wavelength to the peak of the blue component. 
However, if SLSNe II evolution follows closely that of fast-declining type II as suggested by the spectroscopic evolution (see Section~\ref{sec:sp} and Fig.~\ref{fig:lastcmp}), we would not expect any absorption component from the \Ha\/ able to dim the blue peak. Instead an asymmetric, clumpy CSM could explain the difference in intensity between the blue and the red component in \hx\/ and the absence of a red component in \br.  If we considered the blue component of \hx\/ to be a consequence of a boxy profile, and hence more a shoulder than a peak, we would still need to invoke an asymmetric configuration. Such CSM structure would have a two-component wind with a spherically symmetric region responsible for the boxy profile and the presence of a denser clump in the direction opposite to that of the observer in order to explain the intensity of the red component in \hx. Hence both the scenarios would suggest an asymmetric configuration of the CSM. Such configurations have been invoked to explain late time multicomponent \Ha\/ profiles for several type II SNe \citep[e.g.][]{uk86,fe99,pozzo04,in11,ben16}.

\section{Discussion}\label{sec:dis}

\hx\/ and \br\/ provide, for the first time, a small sample of SLSNe II useful for the understanding of the hydrogen-rich SLSNe. These SLSNe are different than those undergoing  strong interaction with an optically thick CSM showing spectra with strong optical lines having multiple emission components such as SN2006gy \citep{smi07}. They have luminosity spanning from $-20>{\rm M}_{g}>-22$, similar spectral evolution and an origin in faint host galaxies. The latter is surprisingly similar to the characteristic locations of their hydrogen-free counterparts \citep{2014ApJ...787..138L,le15,ch15}. 
Their photospheric ($<100$ d) spectroscopic evolution resembles those of bright type II (or type IIL) SNe with a delay of 15-20 days and their light curves show a linear decline over 100 days post peak.  Hence they are more luminous and evolve slower than normal type IIL SNe. These two characteristics (more luminous, slower evolution) 
also link SLSNe I to normal type Ic SNe in a similar way. Therefore, it might be that the power source responsible for the differences between SLSNe I and SNe Ic is also accountable for those between SLSNe II and SNe II.

The extreme luminosity of SLSNe I has been explained with a few scenarios and we use our dataset to constrain similar models for the origin of SLSNe II. 
The light curves that we will use in the following Sections are the full, estimated, bolometric 
light curves that have been corrected for
flux missed in {\em both} UV and NIR as described in Section\,\ref{sec:bol}. 

\begin{figure*}
\includegraphics[width=18cm]{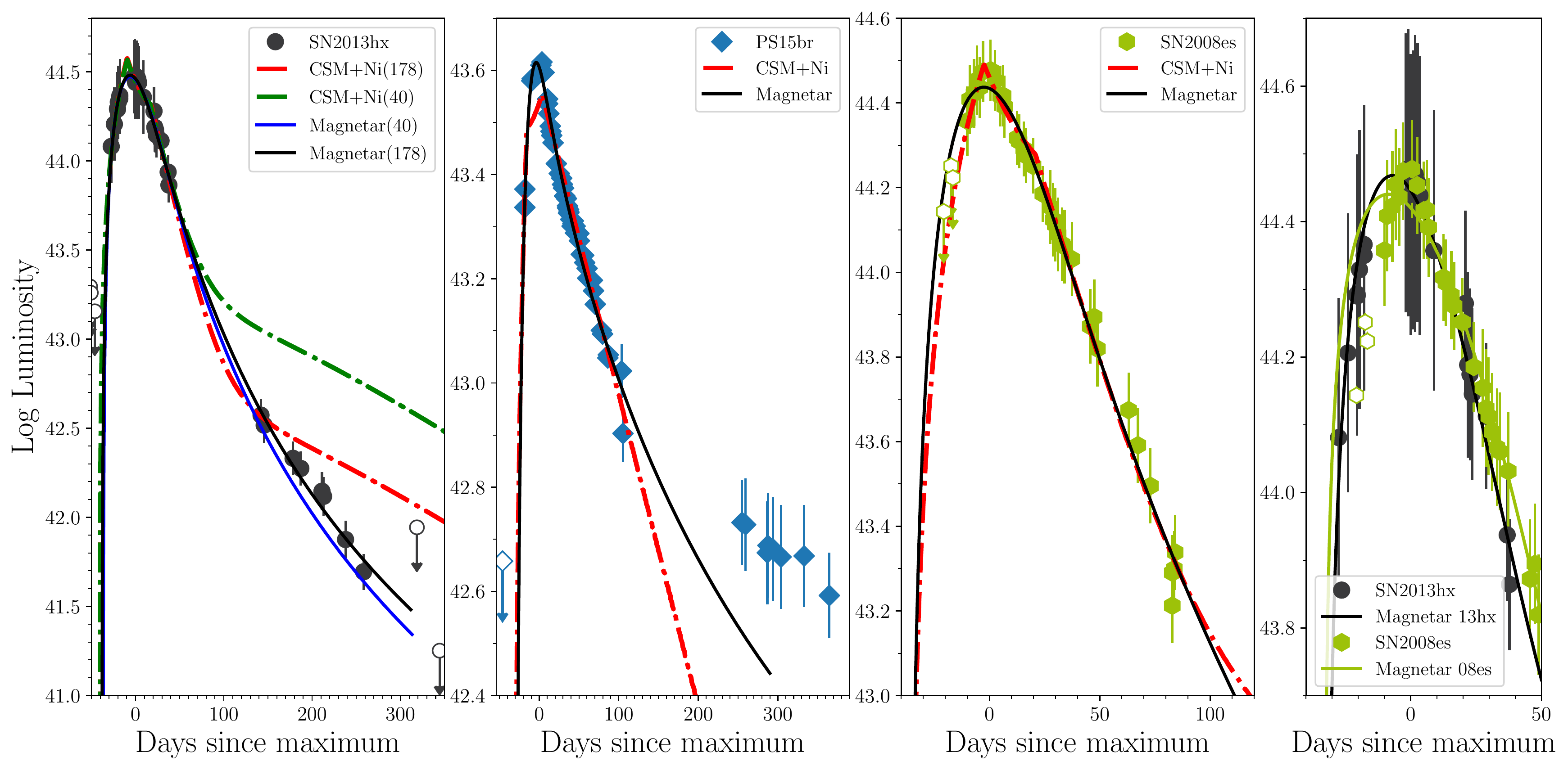}
\caption{Bolometric light curves of \hx\/, \br\/ and \es\/ compared with the best-fit models. The black solid line (and also the blue for \hx) refers to the magnetar model, while the red dot-dashed (and also the green for \hx) refers to the CSM+Ni model. Limits are shown as empty symbols. For the case of \hx, the (40) models consider up to 40 days past maximum, while the (178) refer to the models considering up to 178 days. the plot on the far right shows a comparison of bolometric light curves and magnetar fit around maximum light for SN2013hx and SN2008es. This highlights how the difference time baseline coverage gives slightly different magnetar parameters (see Table~\ref{table:magfit})}
\label{fig:bolfit}
\end{figure*}

\subsection{\ni\/ driven explosion}
To test if radioactive heating could be the main source of energy, we can assume that $\gamma$-rays from \co\/ decay are fully thermalized and can hence estimate  the \ni\/ mass needed for these events comparing their bolometric luminosity with that of SN1987A. 
For SN2013hx, the peak would require a  \ni\/ mass of $M$(\ni\/)$\sim8.0$ \M\/ and the luminosity in the 
tail phase at around 140 days would be $M$(\ni\/)$\sim1.0$ \M.  For PS15br, the luminosity at 100 days would 
require  $M$(\ni\/)$\sim2.2$\M. Although these objects are H-rich and hence  hydrogen recombination should play some role in the photospheric phase of the light curve, we applied our toy model \citep[see Appendix D of][]{in13} to have an alternative estimate of the \ni\/ mass required to power the late light curves.  We 
retrieved values of $4.3 < M$(\ni\/)$ < 8.4$\M\/ for SN2013hx (depending on whether or not we included the tail phase data after 142 days) and  $M$(\ni\/)$\sim4.4$\M\/ for PS15br, for which we do exclude the tail since our first late spectrum showing interaction is coincident with the first photometric point of the tail.

Although there is some uncertainty in these numbers, the large \ni\/ masses could suggest a pair instability explosion \citep{ba67,fr68}. 
However, due to the large ejecta masses expelled in pair instability explosions, the rise times of the light curves is a factor $2-3$ greater than that observed in our objects \citep{wo07}. 
In case of pair instability, the ejecta masses should be $2-5$ times greater than the \ni\/ mass \citep{un08} and such amount of material would be mainly composed of He ($>13$\M) and H ($>11$\M) leading to a long plateau in their light curves \citep{ww95}.
This has not been observed, implying that the late-time evolution of the light curve of \hx\/ and \br\/ is not driven by \ni\/ but is due to some late interaction (see Section~\ref{sec:int}). 

\subsection{Circumstellar medium interaction}
In \hx\/, \br\/ and \es\/ we do not observe narrow and intermediate features suggesting an ongoing interaction with a CSM.
However, we cannot exclude a priori such scenario since
if the CSM is opaque and the shock encounters no further CSM at larger radii, the usual signatures of interaction spectra can be avoided \citep{smm07,ch11}. This could, in principle, explain the spectra evolution in the the objects. 

The dense CSM can be produced by a wind or by a shell ejected prior the SN explosion. In the former case, following the model of \citet{bb93} used to explain type IIL, a dense stellar wind that reprocesses the UV photons produced at shock breakout can boost the luminosity up to superluminous magnitudes. That would need an increase in the mass-loss rate by one to two  orders of magnitude, with respect to those observed in red supergiant (RSG) and hence reaching $\dot{M}\sim10^{-3}$ \M\/ yr$^{-1}$. However, in this scenario after $\sim50-80$ days from peak, the light curve should show a drop in magnitude. The higher the peak luminosity, the greater  the drop with at least a decrease of two mag in a superluminous case - and a subsequent settling onto a \co\/ tail. Our dataset, including \es, does not show such a drop around that phase or even later. 

From the blackbody fit to the SED (see Section~\ref{ss:vel}) and the bolometric luminosity (see Section~\ref{sec:bol}), we can estimate the peak radius of \hx\/ and \br\/ to be $3\times10^{15}$ cm and $2.3\times10^{15}$ cm, respectively.  These radii are a factor 30 and 10 greater than that of a large RSG \citep{smi01} and as a consequence the opaque CSM shell forming the envelope is not bound to the star. Following the formalism of \citet{qu07,smm07}, the peak luminosity produced by the radiation emitted by a shocked, thermalised shell is 
\begin{equation}
L\propto \frac{{\rm M}_{\rm sh}v_{\rm ph}^2}{2\,{\rm t}_{\rm max}},
\end{equation}
where M$_{\rm sh}$ is the mass of the dense CSM shell, $v_{\rm ph}$ is the velocity of the pseudo-photosphere and t$_{\rm max}$ is the rise time. Assuming a rise time of 37 days for \hx\/ and 35 days for \br\/ (cfr. Section~\ref{sec:lc}) and a pseudo photosphere of 6000 \kms\/ (cfr. Section~\ref{ss:vel}) we can infer 2.5~\M\/ and 0.5~\M\/ for the CSM masses of \hx\/ and \br\/, respectively. 

The aforementioned radii imply a wind distance of 300 and 150 AU for the two objects and inferring a wind velocity of 10-100 \kms\/ (which brackets RSG and faster luminous blue variable wind velocities), the wind must have begun $\sim$10-100 yr before the explosion of \hx\/ and $\sim$7-70 yr before that of \br. Hence, the mass-loss rates would be 0.03-0.25 \M\/ yr$^{-1}$ and 0.007-0.07 \M\/ yr$^{-1}$, which are too large to be produced by a 
normal steady stellar wind \citep{smi14}. Therefore if a a dense CSM scenario is invoked as the
explanation of the peak luminosity, it would suggest that  an unbound CSM shell configuration. 

\begin{table*}
\caption{Best-fit parameters for the CSM+Ni model, which has nine free parameters.}
\begin{center}
\begin{tabular}{lccccc}
\hline
\hline
Object & t$_{\rm rise}$ &M$_{\rm CSM}$ & M (\ni) & M$_{\rm ej}$ & $\chi^2$/d.o.f.\\
& (days)  & (\M\/) & (\M\/) & (\M\/) &\\
\hline
\hx*& 40.35/40.35& 2.35/2.37  &2.02/0.61  &9.81/10.48  &0.8/20 1.12/28\\
\br& 31.90&  2.09 & 0.00 &12.94  &500.5/55\\
\es& 36.41 & 2.75  & 1.14  & 9.46 &5.4/30\\
\hline
\end{tabular}
\end{center}
$*$ The first values refer to the model considering only the first 40 days, while the second refer to the model considering up to 178 days.
\label{table:csmfit}
\end{table*}%

\begin{table*}
\caption{Best-fit parameters for magnetar modelling,  which has four free parameters, of the bolometric light curves and $\chi^2$/d.o.f. value, together with the derived parameters (last two columns).}
\begin{center}
\begin{tabular}{lccccc|cc}
\hline
\hline
Object&$\tau_{\rm m}$& $B_{14}$ & $P_{\rm ms}$ & $t_{\rm 0}$& $\chi^2$/d.o.f. &$E^{\rm mag}$ & $M_{\rm ej}$\\
& (day)&  && (MJD) &  &($10^{51}$ erg) & (\M)\\
\hline
\hx\/ *& 36.71/38.15 & 1.31/0.89 & 2.22/2.29 & 56643.20/56641.57&0.6/15 1.61/23& 3.38/5.24 & 6.95/5.20 \\
\br\/      & 17.60  & 1.61 & 6.38 & 57056.27& 161.5/50 & 0.39 & 1.93 \\
\es\/      & 36.58  & 0.97 & 2.35 & 54578.28& 5.0/25 & 2.66 & 5.71 \\
\hline
\end{tabular}
\end{center}
$*$ The first values refer to the model considering only the first 40 days, while the second refer to the model considering up to 178 days.
\label{table:magfit}
\end{table*}

In order to reproduce the bolometric light curve of \hx\/, \br\/ and also \es\/ we use a semi-analytic code based on the equations of \citet{ch12} and implemented by \citet{ni14}. We use the model considering both CSM shell and radioactive decay of \ni\/ contribution to the light curves, setting $\delta=2$ and  $n = 12$ for the SN ejecta inner and outer power-law density profile slopes, respectively, in all cases. We assume Thomson scattering to be the
dominant source of opacity with a solar mixture of mass fraction ($X=0.7$), which gives an opacity $\kappa=0.2(1+X)=0.34$ cm$^{2}$ g$^{-1}$ that is the standard value for a hydrogen and helium rich ejecta. An identical configuration was previously used by  \citet{ch13} to fit the bolometric light curve of \es. Our best fits are shown in Fig.~\ref{fig:bolfit} (red dot-dashed line), while their parameters are reported in Table~\ref{table:csmfit}. We did not fit the tail of \br\/ since that is due to interaction. Since the late \hx\/ tail  may be dominated by interaction - with a different and probably less dense component see Section~\ref{sec:int} - we made two illustrative fits. First, we fitted the light curve using data up to 40 days after maximum and then 
a second using all data out to 178d (before the clear inflection in slope; see Sections~\ref{sec:lc} \&~\ref{sec:bol}) and before the first spectrum showing interaction with another CSM (at 210d, see Section~\ref{sec:sp}). We derive $2.1\lesssim{\rm M}_{\rm CSM}\lesssim2.7\,{\rm M}_{\odot}$, $0.0\lesssim{\rm M}( ^{56}{\rm Ni})\lesssim2.0{\rm M}_{\odot}$ and $9.5\lesssim{\rm M}_{\rm ej}\lesssim12.9{\rm M}_{\odot}$. 
Following the same methodology we also estimate that a wind of 2.6 and 2.3 \M\/ (or 4.7 and 3.9 \M\/ in case of dense shell configuration) are responsible for the late interaction of \hx\/ and \br, respectively.  The \ni\/ masses are consistent with the aforementioned limit and the CSM mass is consistent with observation of a pseudo-photosphere at the velocity reported in Section~\ref{ss:vel}. This mass is of the order of that associated with impulsive mass ejections 
as discussed in \citet{so06}. However, the ejecta masses are at least a factor two greater than what is expected to avoid an optical plateau from hydrogen recombination, while the total masses (ejecta+CSM+remnant) lie in the upper end of those observed and theorised for type II SNe \citep{sma09,jan12,2015PASA...32...16S}. Such CSM masses are more similar to those of stripped-envelope SNe and not of type II. This two component wind CSM configuration that is required to explain the late interaction of \hx\/ and \br, is difficult to explain with a single steady wind. 

\begin{figure}
\includegraphics[width=\columnwidth]{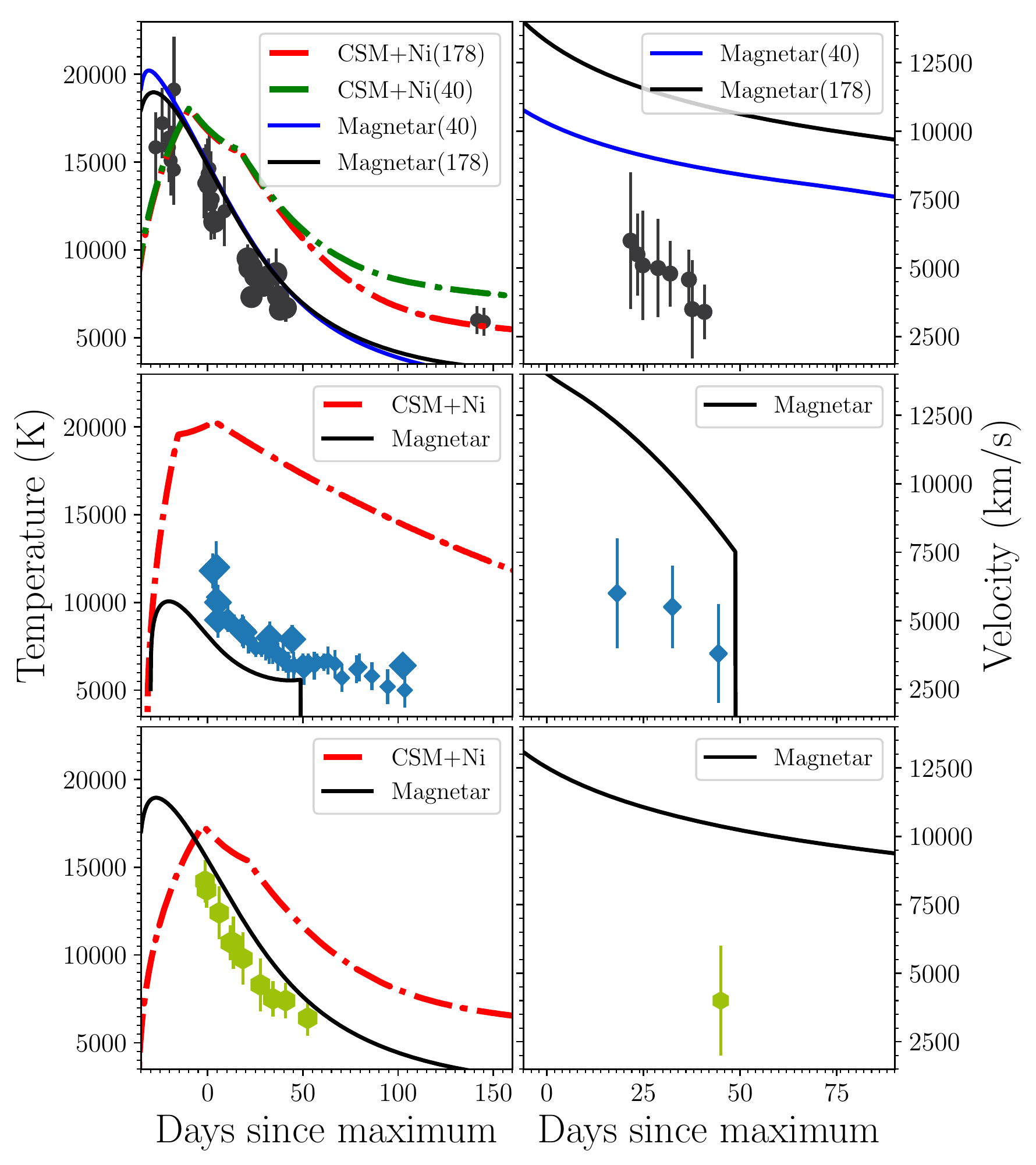}
\caption{Temperature and velocity evolution (only for the Fe~{\sc ii} lines tracing the photospheric evolution) of \hx\/, \br\/ and \es\/ compared with the best-fit models. The black solid line (and also the blue for \hx) refers to the magnetar model, while the red dot-dashed (and also the green for \hx) refers to the CSM+Ni model. Symbols are the same of Fig.~\ref{fig:velTfit}.}
\label{fig:velTfit}
\end{figure}
\subsection{Internal energy source}
Despite the potential of the CSM scenario to describe the main observational data - light curves and spectra evolution - and the overall picture observed in SLSNe II, there are still some details that would be easier to explain with a central engine that deposits its energy into a supernova explosion and significantly enhance the luminosity. 

As observed in Section~\ref{sec:bol}, \br\/ has an asymmetric light curve may be difficult to explain with a CSM interaction, as also shown in Fig.~\ref{fig:bolfit}. In addition SLSNe II have bell shaped light curves around peak only a factor 1.2 wider than those of SLSNe I. Furthermore, they show similar spectra and temperature evolution to those of fast-declining type II SNe but delayed by 15-20 days in a similar fashion to what shown between SLSNe I and stripped envelope SNe. In addition they share similar faint host galaxies as those of SLSNe I.

For these reasons we fit our light curves with our semi-analytical diffusion model presented in \citet{in13} in which a rapidly spinning magnetar deposits its rotational energy into a supernova explosion, through magnetic dipole radiation, and significantly enhances the luminosity \footnote{The semi-analytical code is available at \url{https://star.pst.qub.ac.uk/wiki/doku.php/users/ajerkstrand/start}}. Then, we treat for $\gamma$-ray leakage from ejecta as done in \citet{ch15}, which is similar to the prescription of \citet{wang15}.  The magnetar luminosity depends 
primarily on two parameters, the magnetic field strength B$_{14}$  (expressed in terms of $10^{14}$G) and the initial spin period P$_{\rm ms}$ (in
 milliseconds). As previously done with the interaction model, we used an opacity $\kappa=0.34$ cm$^{2}$ g$^{-1}$.
 Table~\ref{table:magfit} lists the best-fit parameters for each object, and Fig.~\ref{fig:bolfit} shows the fits (black solid line). As the reduced $\chi^2$ ($\chi^2_{\rm red}=\chi^2$/d.o.f.) fitting gives good matches to models {\it without}  \ni\/, we have no need to introduce \ni\/ as an additional free parameter. As in the case of CSM model fitting for \hx, we do two fits (only using data up to 
40 days and then using all data up to 178 days), which in this case are very similar. On the other hand, for \br\/ we only fit the early light curve up to $\sim$102 days. The light curves of our two objects are quite well reproduced but the late luminosity of the model for  \hx\/ is better reproduced by the 40 days magnetar model. These data points
may already have an additional luminosity contribution due to some late interaction (as seen in the late spectra), and hence a light curve dimmer than these points would favour such a model. This is also true for \br\/ where the magnetar model is dimmer than the late time points where the luminosity is driven by interaction. 
The CSM models are too bright to describe \hx\/ data after $\sim$180 days. Furthermore if we remove the \ni\/ contribution, we get a worse $\chi^2_{\rm red}$ and a luminosity at late time still brighter than the data.
The magnetar fit for \br\/ is better than that of the CSM as testified by the $\chi^2_{\rm red}$ values of Tables~\ref{table:csmfit}~\&~\ref{table:magfit}, despite the latter having a higher number of free parameters. 
The magnetar fit to \es\/ is not as satisfactory as the CSM one, as the data show a faster rise than the best fit model. However, the results of \es\/ fit are in agreement with that of \citet{kb10}\footnote{This is expected, since in \citet{in13} we already tested our code with respect to that of \citet{kb10}}.
One may ask why \hx\/ and \es\/ end up with different results if their light curve behaviour is so similar. 
As shown in the right panel of Fig.~\ref{fig:bolfit}, the reason is that we have different time coverage for each. The light curve data for \es\/ stretch to 80\,days 
after peak where as the data for \hx\/ stops at 50 days, and there are more pre-peak detection 
points for SN2013hx than for \es\/ (which has only limits before -10 days). 

We further estimated velocity and temperature evolution from the light curve fits to ejecta mass and kinetic energy.
In Fig.~\ref{fig:velTfit} we show that the temperature evolution of the magnetar model matches the observed ones reasonably well and better than the interaction model, which produces results which are always too hot.  Photospheric velocities are slightly faster than the Fe~{\sc ii} velocities, which trace the photospheric evolution. The sudden drop to zero of both temperature and velocity evolution of the best fit model for \br\/  happens because more than half of the core has been exposed. This is a consequence of the low energy and ejecta mass needed to fit the light curve. 
 
Another possible central engine is that of the fall-back accretion \citep{dk13}, which can give a similar asymptotic behaviour of the light curve (L$_{\rm t}\propto t^{-5/3}$) with respect to the magnetar engine but would have lower luminosity than that observed after 180 days in \hx. However, this scenario implies photospheric velocities higher than that shown by our sample and would fail to reproduce \br\/ light curve since models considering H recombination have longer rise times and a bump 200 or 300 days after the explosion. To better investigate this scenario a dedicated fit for these objects is required.

\section{Conclusion}\label{sec:end}

We have presented extensive photometric and spectroscopic coverage of two superluminous, hydrogen-rich supernovae with light curves covering from weeks before peak up to 260 (\hx) and 365 (\br) days.  
We then analysed the two objects together with the only other SLSN II (\es) to search for similar observational behaviours suggesting their existence as superluminous events different than superluminous interacting event (i.e. SLSNe IIn) or SLSNe I.

\hx\/ has M$_g = -21.7$ mag and $\mathrm {L_{\rm bolometric}}\approx2.75\times10^{44}$ erg s$^{-1}$ at peak, while \br\/ shows M$_g \sim -20.2$ and M$_U \leq -20.8$ mag and $\mathrm {L_{\rm bolometric}}\approx4.15\times10^{43}$ erg s$^{-1}$ at maximum light. The former has a post peak light curve decline similar to that of \es\/ and normal type IIL, whereas \br\/ shows an initial decline comparable to that of \hx\/ and \es/ and a second, slower decline similar to those shown by transitional type IIn/II events at normal luminosity. Although \br\/ luminosity is fainter than \hx\/ and \es\/, we have consider it a ``bona fide'' SLSN II due to its spectroscopic evolution similar to that of \hx\/ and \br\/ and a spectrophotometric evolution dissimilar to any other H-rich SN in a similar luminosity space.

For \br\/ we also obtained an epoch of broadband polarimetry at 6 days after maximum. After correcting for the ISP and the polarisation bias, we retrieved a polarised value of $P = 0.94 \pm 0.17$\%. However, this value strongly depends on the assumed ISP and hence no conclusive evidence on the source of the polarisation associated with \br\/ can be derived.

SLSNe II (namely \es, \hx, \br) show a spectroscopic evolution dominated by Balmer, He, Ca, Fe and other metal lines and an emission feature at early times at $\sim$4600~\AA\/ identified as C~{\sc iii}/N~{\sc iii}.
\hx\/ and \br\/ show a late interaction (phase$>$140d and $>$250d, respectively) due to an asymmetric, clumpy CSM that dominates the late light curve behaviour. 
Regardless of the peak luminosity, SLSNe II spectroscopic and temperature evolution resembles those of fast-declining type II (or type IIL) SNe with a delay of 15-20 days. This suggest as the transients are linked to type II SNe in a similar fashion as SLSNe I are linked to type Ic SNe.

We applied two semi-analytical codes to fit the light curves
of all SLSNe II. The first based on the interaction between the ejecta and a dense uniform shell of H-rich material, while the second based on a diffusion model with energy input
from a spinning down magnetar.  
We were able to reproduce the available bolometric light curve with 
$2.1\lesssim {\rm M}_{\rm CSM}$(\M)~$\lesssim2.8$ and 
$9.5\lesssim {\rm M}_{\rm ej}$(\M)~$\lesssim13.0$. 
While the late time light curve exhibited by \hx\/ and \br\/ and driven by interaction can be fitted with a wind of 2.6 and 2.3 \M\/ or a dense shell of 4.7 and 3.9 \M, respectively
All light curves are also reproduced with feasible physical values for a magnetar including the data at 140-180 days of \hx\/, when the luminosity of our best fit model is lower than the data, which are already due to the aforementioned late interaction. To fit them we require $0.9\lesssim B_{14}$
$\lesssim1.6$, $2.2\lesssim P_{\rm ms}$$\lesssim6.4$ consistent with
$B$ of known galactic magnetars ($B_{14}\sim1-10$) and with
physically plausible periods ($P_{\rm ms}>1$). We derived energies of
$0.4\lesssim E^{\rm mag}$($10^{51}$ erg) $\lesssim5.2$ and ejected masses of
$1.9\lesssim M_{\rm ej}$(\M)~$\lesssim 6.9$.
Both scenarios have their weaknesses. The high ejecta masses retrieved by the interaction model and the spectroscopic similarities to bright, linear type II regardless of the peak luminosity, would tend to disfavour of the interaction scenario. On the other hand, these characteristics can be better explained by the magnetar scenario. However, at least for the case of \br\/, some weak interaction is present for the majority of the photospheric evolution in a similar fashion to hat found for same SLSN I \citep{in17}.
In addition, the interaction shown at late phase by \hx\/ and \br\/ does not allow a firm conclusion and suggests as interaction could play a role in the evolution of SLSNe II.

Despite our dataset and analysis of this first sample of SLSNe II, open questions remain 
\begin{itemize}
\item Do SLSNe II with interaction-free light curves exist? What should be their luminosity and what are
the line profiles expected in a truly nebular spectrum?
\item What is the role of metallicity in the progenitor stars
  evolution that will produce a H-rich SLSNe? It appears that they
  are all associated with faint dwarf galaxies and possibly
  low metallicity progenitors as those of SLSNe I. 
\item Can the geometry of the explosion play a role in the observables and/or shed light on the powering mechanism or their light curves?
\end{itemize}

As for the case of SLSNe I, to address these topics, high quality data and their modelling at early phase or in an interaction-free nebular phase are needed
to determine the ejecta masses, composition and the mass of 
\co\/ contributing to the luminosity. This seems the most likely way to make
progress.

\section*{Acknowledgments}
SJS acknowledges funding from the European Research Council under the European Union's Seventh Framework Programme (FP7/2007-2013)/ERC Grant agreement n$^{\rm o}$ [291222] and  STFC grants ST/I001123/1 and ST/L000709/1. MF is supported by a Royal Society - Science Foundation Ireland University Research Fellowship
EEEG acknowledges support for this work by the Deutsche Forschungsgemeinschaft through the TransRegio project TRR33 'The Dark Universe'.
SB is partially supported by the PRIN-INAF 2014 project Transient Universe: unveiling new types of stellar explosions with PESSTO.
Support for I.A. was provided by NASA through the Einstein Fellowship Program, grant PF6-170148.
KM acknowledges support from the STFC through an Ernest Rutherford Fellowship.
SP and RC acknowledge support from EU/FP7-ERC grant no [615929].
SS acknowledges support from CONICYT-Chile FONDECYT 3140534, Basal-CATA PFB-06/2007,
and Project IC120009 ``Millennium Institute of Astrophysics (MAS)'' of Iniciativa
Cient\'{\i}fica Milenio del Ministerio de Econom\'{\i}a, Fomento y Turismo.
This work was partly supported by the European Union FP7 programme through ERC grant number 320360.
Part of this research was conducted by the Australian Research Council
Centre of Excellence for All-sky Astrophysics (CAASTRO), through project
number CE110001020
Pan-STARRS is supported by the University of Hawaii and the National Aeronautics and Space Administration's Planetary Defense Office  under Grant No. NNX14AM74G.  The Pan-STARRS1 Sky Surveys have been made possible through contributions by the Institute for Astronomy, the University of Hawaii, the Pan-STARRS Project Office, the Max Planck Society and its participating institutes, the Max Planck Institute for Astronomy, Heidelberg and the Max Planck Institute for Extraterrestrial Physics, Garching, The Johns Hopkins University, Durham University, the University of Edinburgh, the Queen's University Belfast, the Harvard-Smithsonian Center for Astrophysics, the Las Cumbres Observatory Global Telescope Network Incorporated, the National Central University of Taiwan, the Space Telescope Science Institute, and the National Aeronautics and Space Administration under Grant No. NNX08AR22G issued through the Planetary Science Division of the NASA Science Mission Directorate, the National Science Foundation Grant No. AST-1238877, the University of Maryland, Eotvos Lorand University (ELTE), and the Los Alamos National Laboratory. This work is based (in part) on observations collected at the European Organisation for Astronomical Research in the Southern Hemisphere, Chile as part of PESSTO, (the Public ESO Spectroscopic Survey for Transient Objects Survey) ESO programs188.D-3003, 191.D-0935. 
Some of the data presented herein were obtained at the Palomar Observatory, California Institute of Technology.
The Liverpool Telescope is operated on the island of La Palma by Liverpool John Moores University in the Spanish Observatorio del Roque de los Muchachos of the Instituto de Astrofisica de Canarias with financial support from the UK Science and Technology Facilities Council.

\bibliographystyle{mnras}
\bibliography{cosmo.bib}\label{bib}

\clearpage

\appendix
\section{Observations and Data reduction}\label{sec:data}

Both objects were immediately selected by PESSTO as follow-up science targets and 
a combination of optical, near infrared (NIR) and ultraviolet (UV) photometric monitoring 
was carried out. The epochs of all the data, as well as the telescopes and instruments used are listed in Appendix~\ref{sec:tab}. 

All images were reduced (trimmed, bias subtracted and flat-fielded) by the SMT, PSST, Las Cumbres Observatory  \citep[LCO,][]{lcogt}, Liverpool Telescope \citep[LT,][]{2004SPIE.5489..679S}, CRTS and PESSTO pipelines.
Photometric zero-points and colour terms were computed using observations of standard fields ($U,B,V,R,I$ in Vega and $g,r,i,z$ in AB system). We then calibrated the magnitudes of local stellar sequences shown in Appendix~\ref{sec:ss}. The average magnitudes of the local-sequence stars were used to calibrate the photometric zero-points in non-photometric nights.  
The near-infrared $J,H,K$ photometry was carried out on SOFI on the NTT and the data 
were again reduced as discussed in \citet{pessto}. Magnitudes were 
calibrated to the Two Micron All Sky Survey system using  local sequence stars. The magnitudes of the SNe, obtained through a point spread function (PSF) fitting technique using standard procedures in IRAF\footnote{Image Reduction and Analysis Facility, distributed by the National Optical Astronomy Observatories, which are operated by the Association of Universities for Research in Astronomy, Inc, under contract to the National Science Foundation.}, were measured on the final images. The uncertainties reported in the Appendix~\ref{sec:tab} Tables~\ref{table:sn13hx},~\ref{table:sn15br}~\&~\ref{table:sn15br2} were estimated by combining in quadrature the errors of photometric calibration and those on the instrumental magnitudes.
When the objects were not detected, limiting magnitudes were estimated by placing artificial stars of different magnitudes at the expected SN positions. When necessary - i.e. after 170d for \hx\/ and 100d for \br\/ - we applied a template subtraction technique
\citep[through the HOTPANTS\footnote{http://www.astro.washington.edu/users/becker/hotpants.html} package based on the algorithm presented in][]{al}. The instruments used to obtain the templates were NTT for \hx\/ and PSST for \br. The same frames were used to measure the host magnitudes and host NIR limit magnitudes (see Sections~\ref{sec:hx}~\&~\ref{sec:br}).

Differences between passbands were taken into account applying a passband correction (P-correction) using the SuperNova Algorithm for P-correction (SNAP) in the S3 package (see Appendix~\ref{sec:s3} for further details). This P-correction is similar to the S-correction \citep{str02,pi04}.

Ultraviolet \citep[$uvw2, uvm2, uvw1$; see][]{poo08} observations, obtained by UVOT on board of the {\it Swift} satellite (P.I. Margutti for \hx\/ and Inserra for \br) were reduced using the HEASARC\footnote{NASA's High Energy Astrophysics Science Archive Research Center} software. We
analysed these publicly available data independently. {\it Swift} $u, b, v$ aperture magnitudes were transformed to Johnson system by applying a shift after comparison with optical ground-based data taken close in time \citep[see][for further details in the procedure]{in11,pasto13}.

Host galaxies photometry was carried out through aperture photometry within {\sc iraf/daophot} package, and
we used the same aperture size to measure the flux of local secondary standards in the field. The $griz$ photometry
was calibrated against Pan-STARRS1 sequence stars, or local secondary stars, and converted to the SDSS photometric system; while JHK was calibrated against the 2MASS catalogue.

The journal of spectroscopic observations is listed in Table\,\ref{table:sp}. The majority of the follow-up spectra where taken 
with PESSTO and the NTT. 
Integral field spectra were also taken with the ANU 2.3 m telescope + the Wide Field Spectrograph \citep[WiFeS,][]{wifes} at Siding Spring Observatory in New South Wales, Australia.  A final spectrum of \br\/ was taken with the SuperNova Integral Field Spectrograph \citep[SNIFS,][]{la} at the 2.2m University Hawaii Telescope.
A series of spectra of \hx\/ was taken with the  VLT+FORS2\footnote{Very Large Telescope + FOcal Reducer and low dispersion Spectrograph} in the second season of observations, and combined together into one deep late-time spectrum. These FORS2 exposures had a total of 11$\times$1230 sec and 1$\times$1800  and were
taken on the nights of 22-24 October 2014 through service mode on the VLT (giving an on-sky total
exposure time of  4hr\,15min\,30sec). On the other hand, two spectra of \br\/ were taken with the VLT+XShooter in service mode. We used the custom made pipeline described in \cite{2015A&A...581A.125K} which takes the ESO pipeline produced 2D spectral products 
\citep[from {\sc reflex};][]{freud13} and uses  optimal extraction with a Moffat profile fit.  This pipeline produced flux calibrated spectra with rebinned dispersions of 0.4 \AA\,pix$^{-1}$ in the UVB+VIS arms and  0.6 \AA\,pix$^{-1}$ in the NIR arm.  The 
NTT spectral data  were reduced using the PESSTO pipeline \citep{pessto}, while the VLT+FORS2
spectra were reduced in the {\sc reflex} environment \citep{freud13} and the SNIFS spectrum using standard IRAF procedures as described in \cite{smartt16b}. 
Optimal extraction of the spectra was adopted to improve the final signal-to-noise (S/N) ratio. Wavelength calibration was performed
using spectra of comparison lamps acquired with the same configurations as the SN observations. Atmospheric extinction correction was based on tabulated extinction coefficients for each telescope site. 
The WiFeS spectra were reduced with {\sc pywifes}\footnote{http://www.mso.anu.edu.au/pywifes/} package \citep{Chi14} to produce data cubes, from which the final spectra were obtained using a PSF weighted extraction routine.
Flux calibration was performed using spectro-photometric standard stars observed on the same nights with the same set-up as the SLSNe. The flux calibration was checked by comparison with the photometry, integrating the spectral flux transmitted by standard {\it griz} filters and adjusted by a multiplicative factor when necessary. The resulting flux calibration is accurate to within $0.1-0.2$ mag.

All reduced spectra (and calibrated NIR images), taken at the NTT before 2014 April 30, are available from the ESO Science Archive Facility as PESSTO SSDR2 and details of data access are provided on the PESSTO website\footnote{www.pessto.org}. Data taken after this date are part of the SSDR3. All spectra will also be
available through WISeREP\footnote{http://wiserep.weizmann.ac.il/home} \citep{wise}. 

\section{S3 package}\label{sec:s3}
The S3 package is a python suite publicly available at \url{https://github.com/cinserra} and accepts fits, txt, dat and ascii files as input. It contains the following programs.

\subsection{snake}
The SuperNova Algorithm for $K$-correction Evaluation ({\sc {\sc snake}}) and its light version developed to handle multiple input files in a less interactive fashion ({\sc {\sc snake}LOOP}) allow to evaluate the $K$-correction from the observed passband (P) to the rest-frame, desired passband (R) according to the formula defined in \citet{ho02,bl07}:

\begin{equation}
\begin{split}
K_{\rm PR} = -2.5\ {\rm log}\left[ \frac{1}{1+z} \right]\ &\\
-2.5\ {\rm log}\left[\frac{\int d\lambda_{\rm o}\: \lambda_{\rm o}\: L_{\lambda}(\lambda_{\rm o}/(1+z))\: P(\lambda_{\rm o}) \int d\lambda_{\rm e}\: \lambda_{\rm e}\: g_{\lambda}^{R}(\lambda_{\rm e})\: R(\lambda_{\rm e})}{\int d\lambda_{\rm o}\: \lambda_{\rm o}\:  g_{\lambda}^{P}(\lambda_{\rm o})\: P(\lambda_{\rm o}) \int d\lambda_{\rm e}\: \lambda_{\rm e}\:L_{\lambda}(\lambda_{\rm e}) R(\lambda_{\rm e})}\right]\ , &
\end{split}
\end{equation}

where P($\lambda$) and R($\lambda$) are the filter response per unit photon, g$_\lambda^{\rm passband}$ are the flux densities per unit wavelength  for the standard source for P and R, $\lambda_{\rm o}$ refers to the observer frame and $\lambda_{\rm e}$ to the rest-frame. We note that the S3 package uses flux and not density flux, hence the zero points are also treated accordingly. GALEX FUV and NUV; UVOT $uvw2$, $uvm2$, $uvw1$; Johnson $UBVRI$; Sloan $u,g,r,i,z$; 2MASS $J,H,K$; Euclid NIR $Y,J,H$ \citep{in17b} and 4000 \AA\/ and 5200 \AA\/ passbands \citep[created for the use of SLSNe as cosmological probes in][]{in14} are recognised by the program and the cross $K$-correction - which is when the observed and rest-frame filter shapes are the most similar - is suggested. In the case that more than 50~\AA\/ of one of the two filters chosen are not covered by the observer frame or rest-frame spectrum, the program allows the user to combine the spectrum with the best black-body fit to the data.
The output of the program is the $K$-correction ($K_{\rm PR}$) and it is related to the apparent and absolute magnitude as follows:

\begin{equation}\label{eq:sti}
{\rm M_{\rm R}}= {\rm m}_{\rm P} -(5\:{\rm log}\: {\rm D_{\rm L}+25)  - A_{P}} - K_{\rm PR}\ ,
\end{equation}

in which M$_{\rm R}$ is the absolute magnitude in the rest-frame filter, m$_{\rm P}$ is the apparent magnitude in the observed filter, D$_{\rm L}$ is the distance luminosity of the source, A$_{\rm P}$ is the foreground reddening extinction toward the source in the observed band.
Assuming as a first approximation that the terms are uncorrelated and evaluating the single term errors as their variance, the overall error on the $K$-correction is a root mean square:

\begin{equation}\label{eq:err}
\sigma_{K} = \sqrt{\frac{\sigma^2_{z} + \sigma^2_{{\rm ZP}_{\lambda_{\rm o}}}+  \sigma^2_{{\rm ZP}_{\lambda_{\rm e}}}+\sigma^2_{\rm BB} }{\rm N}},
\end{equation}

where $\sigma_{z}$ is the error on the redshift, which is set by default to be $\pm0.005$ and can be changed by the user; $\sigma_{ZP_{\lambda}}$ are the errors on the zero points for the observed (o) and rest-frame (e) filters evaluated from the average of all the differences between the zero points of the standard passbands (in the Vega and AB system) and those of the filters used by the telescopes listed in SNAP (cfr. Section~\ref{sec:snap}); $\sigma_{\rm BB}$ is the error on the assumption that a blackbody can be used to measure the wavelength uncovered by the spectrum for the filter chosen; N is the number of errors considered. Other two errors - related to the spectra template used in the case a spectrum of the object for the same, or very close in time, epoch of the photometry are not available - should be considered, but the program does not handle that since they are highly dependent to the nature and the evolutionary phase of the object for which the $K$-correction is needed.

In order to test the program reliability we compared our results with those published by \citet{kim96} finding similar values. A previous, beta version, based on IRAF and STDAS was used to calculate $K$-corrections in \citet{in14,in15}, while this version has been used by \citet{gall15} and \citet{pol16}. Furthermore, the program can also evaluate the $K$-correction when reddening is also applied and hence directly evaluate the term $-({\rm A_{\rm P}} + K_{\rm PR})$ of equation~\ref{eq:sti}. However, this is not the correct way and the additional errors related to the assumed reddening and reddening law are not accounted for.

\subsection{SNAP}\label{sec:snap}
The SuperNova Algorithm for P-correction (SNAP) evaluate a P-correction given by:
\begin{equation}
{\rm P}_\lambda = {\rm F}_\lambda \times {\rm QE}_\lambda \,
\end{equation}

where F$_\lambda$ is the filter transmission function and QE$_\lambda$ is the quantum efficiency of the detector. In contrast with the S-correction, the program does not take in account the lens throughput since, despite the use of a few materials, they are relatively flat across the optical range. However, this might cause some problems at wavelengths bluer than 3400 \AA, i.e. affecting the $U$ and $u$ passbands. The atmospheric transmission profile is also ignored since SN magnitudes are usually evaluated through sequence stars 
calibrated with Sloan stars (and hence such correction is already taken in account at this level) or with Landolt stars and subsequent use of 
programs that apply such correction. We tested a few mirror reflectivity functions and found small dissimilarities between them, quantifiable in magnitudes differences smaller or comparable with the usual photometric errors; hence we decide to not include such additional term.
Then the P$_\lambda$ allows to correct the apparent magnitude of a given telescope to that of the standard passband system as follows:

\begin{equation}
{\rm m_P} = {\rm m_F} - {\rm P}_\lambda,
\end{equation}

where m$_{\rm P}$ is the apparent magnitude with the standard passbands (Johnson $U,B,V,R,I$ or Sloan $u,g,r,i,z$) and m$_{\rm F}$ is the one measured with photometry at a given telescope. As {\sc {\sc snake}} also {\sc SNAP} evaluate the errors of the P-correction in the same way of equation~\ref{eq:err} but without the redshift term and with the rest-frame filter considered as the error on the zero point in flux of the standard passband.

We retrieved the information needed for the P-correction for the New technology Telescope (NTT; ESO website), La Silla-Quest \citep[LSQ;][and D. Rabinowitz private communication]{bal07}, SkyMapper \citep[SMT;][M. Renault and R. Scalzo private communications]{bes11}, the Liverpool Telescope (LT; LT website), the Las Cumbres Observatory Global Network (LCOGT; S. Valenti private communication), the Optical Gravitational Lensing Experiment IV \citep[OGLE-IV;][and L. Wyrzykowski private communication]{ud15}, the North Optical Telescope (NOT; NOT website), the Pan-STARSS1 survey \citep[PS1;][and K. W. Smith private communications]{to12}, the Telescopio Nazionale Galileo (TNG, TNG website), the Copernico Telescope at Cima Ekar in Asiago (EKAR, S. Benetti private communication).

\subsection{SMS}
The Synthetic Magnitudes from Spectra (SMS) program is just a python version of the STSDAS/HST IRAF tool calcphot. It evaluates the flux in a given passband and converts it in magnitudes (AB or Vega system) for the same filter list used by {\sc {\sc snake}}. As {\sc {\sc snake}}, it evaluates the errors in the same fashion of {\sc {\sc snake}} but without the terms about the redshift and the rest-frame filter

\section{Tables}\label{sec:tab}

\begin{table*}
\caption{$U,B,V,g,r,i,J,H,K$ magnitudes of \hx\/ and assigned errors in brackets.}
\begin{center}
\scriptsize
\begin{tabular}{cccccccccl}
\hline
\hline
Date & MJD & Phase*  &$ U$ &$B$ & $V$ & $g$ & $r$ & $i$ & Inst.\\
dd/mm/yy &  & (days) & & && & &\\
\hline
01/10/13 &   56566.61   & -104.32  &-&   - &	   -	  &  $>$20.86 &   $>$20.68 &    -     &     SMT\\
08/10/13  &  56573.58   & -98.16 &- &  -  &	-   	  &      -  &   $>$20.71    &   -    &     SMT\\
19/10/13   & 56584.62  & -88.39  & -  & - &	  -   &	  $>$18.97     &  $>$19.24       &    -    &    SMT\\
28/10/13   & 56593.63  & -80.41  & - &-   &	  -   &	  $>$19.89     &  $>$19.65       &    -      &  SMT\\
01/11/13   & 56597.64   & -76.87  & - & - &	  - 	  &  $>$20.19   &  $>$19.98    &    -   &    SMT\\
06/11/13   & 56602.60   & -72.48  & - & - &	-     &	  $>$20.24     & $>$20.02      &  -    &     SMT\\
10/11/13   & 56606.64   & -68.90 &-   & - &	-     &	  $>$19.63     & $>$19.64      &   -    &     SMT\\
15/11/13   & 56611.65   & -64.47 &-  &  - &	-     &	  $>$19.20    & -    &      -   &    SMT\\
20/11/13   & 56616.62  &  -60.07  & - & - &	-     &	  $>$18.59    &  $>$18.68    &    -     &     SMT\\
25/11/13   & 56621.60   & -55.66 & -  & - &	-     &	  $>$20.03   &   $>$19.95     &   -     &     SMT\\
30/11/13   & 56626.58  & -51.26  & - &  - &	-     &	  $>$19.94   &   $>$19.91     &    -   &    SMT\\
01/12/13   & 56627.55  & -50.40  & -  &-  &	-     &	  $>$20.30    &  $>$19.99      &  -     &    SMT\\
06/12/13   & 56632.55  &  -45.97 & -  &-  &	-     &	   -     &   $>$20.28    &    -   &     SMT\\
27/12/13   & 56653.51  & -27.42  &  -&-   &	-   	  &  17.86   (0.04)    & 18.14   (0.06)  &     -    &    SMT\\
31/12/13   & 56657.55  & -23.85  & -  &-  &	-   	   & 17.50   (0.06)  &   17.83   (0.08)    &   -    &   SMT\\
04/01/14   & 56661.48  & -20.37  & -  &-  &	-  	  &  17.33   (0.02)   &  17.61   (0.06)    &  -   &      SMT\\
04/01/14   & 56661.49  & -20.36  & -  & - &	-     &	  17.32   (0.02)  &   17.62   (0.06)   &    -   &     SMT\\
05/01/14   & 56662.50  &  -19.47 & -  &-  &	-     &	  17.27   (0.04)  &   17.52   (0.05)   &   -     &    SMT\\
07/01/14   & 56664.49  & -17.71  &  - & -  &	-     &	  17.09   (0.26)  &   17.47   (0.08)   &   -      &   SMT\\
07/01/14   & 56664.50  &  -17.70 & -  &-  &	-     &	  17.20   (0.05)  &   17.43   (0.08)  &    -     &    SMT\\
25/01/14   & 56682.47  & -1.80  &  - &  - &	   -	  &  16.88   (0.04)    & 17.07   (0.08)    &   -   &     SMT\\
26/01/14   & 56683.51  & -0.88 & -  &  - &	  -   &	  16.89   (0.03)    & 17.09   (0.08)    &    -  &     SMT\\
26/01/14   & 56683.52  & -0.87 &-   & -  &	   -	  &  16.88   (0.03)    & 17.06   (0.09)  &    -     &    SMT\\
27/01/14   & 56684.50  & 0.00 &  -  & - &	  - 	  &  16.86   (0.27)  &   17.12   (0.13)  &    -     &    SMT \\
28/01/14   & 56685.49  & 0.88 &-   & -  &	 -    &	  16.89   (0.05)   &  17.12   (0.11)  &    -     &    SMT\\
29/01/14   & 56686.48  & 1.75 &-   & -  &	 -  	  &  16.94   (0.04)     &17.06   (0.10)  &       -    &  SMT\\
29/01/14   & 56686.49  & 1.76 &-   &-   &	-     &	  16.95   (0.05)   &  17.13   (0.11)  &     -     &   SMT\\
30/01/14   & 56687.48  & 2.64 & -  &-   &	-     &	  16.95   (0.06)   &  17.09   (0.09)   &    -     &   SMT\\
31/01/14   & 56688.47  & 3.51 &  - &-   &	-     &	-       &  17.14   (0.11)  &      -    &   SMT\\
06/02/14   & 56694.49  & 8.84 &  -  &	-  &  -   &	  17.06   (0.05)  &   17.16   (0.09)   &    -    &    SMT\\
19/02/14   & 56708.03  & 20.82 &   - &  -  &  16.76  (0.02)   &   -   &  -    &   -      &   NTT\\
20/02/14   & 56709.02  & 21.70  &  - &  -  &  16.83  (0.01)  &  -     &  -   &     -     &   NTT\\
21/02/14  & 56709.88 &22.56& 17.08 (0.09)  &17.67 (0.09) & 16.89 (0.12) &  -  & -&   - &    SWIFT\\
21/02/14   & 56710.02  & 22.58  & -   & - &   16.86  (0.02)  &    -    &   - &     -   &     NTT\\
21/02/14  &  56711.02  & 23.47 & -  &   -&    16.87   (0.02)   &   -  &   -    &      -   &   NTT\\
24/02/14  & 56712.82   &25.17 &17.35 (0.08)  & 17.74 (0.10)  & 17.04  (0.19)  &  - & - &   - &   SWIFT\\
27/02/14   &56715.36  &27.43 & 17.48 (0.08)   &17.91 (0.10)  & 17.21  (0.31)  &  -& -   &    -&   SWIFT\\
28/02/14  &  56717.04  & 28.80 &-   & -  &    17.06  (0.02)    & -   &  -    &     -    &    NTT\\
03/03/14 & 56719.16  &30.81&17.73 (0.08) & 17.78 (0.10)  &17.15  (0.14)   &- &  - &   -  & SWIFT\\
05/03/14 & 56721.17 & 32.59&17.74 (0.09) & 17.92 (0.10)  &17.22  (0.14)   &-  & -   &   -   & SWIFT\\
08/03/14 & 56724.56 & 35.61&17.96 (0.09) & 18.07 (0.10)&  17.29  (0.12)  & - & -  &   -   & SWIFT\\
08/03/14  &  56725.02  & 35.86  & -   & -   &  17.28  (0.02)  &   -   &   -   &     -     &   NTT\\
09/03/14  &  56726.02  & 36.74 &  -  & -   &  17.40   (0.02)   & -      &  -   &   -       &  NTT\\
10/03/14  &  56727.02  & 37.62  & -   &  -  &  17.52  (0.02)   & -      &  -   &   -     &    NTT\\
06/07/14  &  56844.40  & 141.50 & - & 22.30  (0.15)   &  21.80   (0.10)    &  22.32   (0.10)    & 21.80    (0.10)   &   20.70  (0.10)    &   LCO\\
10/07/14  &  56848.40  & 145.04 & - & 22.48  (0.15)   &  22.05   (0.10)    &  22.49   (0.10)    & 21.88    (0.10)   &   20.83  (0.10)    &   LCO\\
13/07/14  &  56851.72  & 147.98  &   -  &  -  &   -     &  -     & $>$21.50     &    -    &   LCO\\
31/07/14  &  56869.68  & 163.88  & -     & - &    -   &	  $>$21.29  &  $>$20.94  &      -  &   SMT\\
15/08/14  &  56885.23  &  177.64 & -   & -&     -    & -   &      21.74 (0.09)    & -   &    NTT\\
25/08/14  &  56895.17  & 186.43   &  - &-  &    -   &   -   &      21.87$^{\dagger}$  (0.06)   & 21.50  (0.05)   &  NTT\\
21/09/14  &  56922.10  & 210.27  & -  &-  &    23.26 (0.19)  &   -   &       22.34$^{\dagger}$  (0.06)   &     -  &   NTT\\
23/09/14  &  56924.15  & 212.08  & -  & - &   -    &  23.72 (0.10)  &  22.54  (0.14)   & 21.80  (0.09)   &  NTT\\
21/10/14  &  56952.15  & 236.86  &-  &  - &    -    & 24.75 (0.17)  & 23.23  (0.24)   & 22.35  (0.14)   &  NTT \\
22/12/14  &  56975.11  & 257.18  &  -  & - & -    &  24.96 (0.26)    &23.68 (0.29)   & 22.80  (0.19)   &  NTT \\
18/02/15  &  57072.04  & 344.48  &-   &  - &   -  &    -  &  -& $>$24.00  &  NTT \\
Host A&   & &  &  &  &   &  & \\
20/01/15  &  57043.09  & 318.75  &  -   &- &     -&   -   &23.26 (0.16) & -  &  NTT \\
10/12/15 & 57367.12 &   606.77  &  -  & -&  -   &   24.43 (0.16)   & 23.20 (0.16)  &  21.82 (0.16) &  NTT \\
Host B&   & &  &  &  &   &  & \\
01/02/16 & 57420.08 &   653.85  &  -  & -&  -   &   24.71 (0.38)   & 24.55 (0.35)  &  23.54 (0.32) & Magellan \\
\hline
\hline
Date & MJD & Phase* & & $J$ & $H$ & $K$ & & & Telescope  \\
dd/mm/yy &  & (days)  &  & &\\
\hline
23/11/14 &  56985.12  & 266.03 && 	 22.21  (0.18) &    $>$22.50 & 	  21.48 (0.12)& & &NTT\\
20/12/14  & 57012.10 &  289.91 && 	  -    &  -	&   22.10 (0.42) &   && NTT\\  
29/12/14  & 57021.02  & 297.90 & & -  & -  &$>$22.18  & &   & NTT\\
Host A &   & &  &  &  &   &  & \\
17/12/15  & 57374.12 & 613.00& &   $>$23.10&  $>$23.00 &  $>23.20$& &   & NTT\\
\hline
\end{tabular}
\end{center}
* Phase with respect to the $r$-band maximum.\\
${\dagger}$ $r$ in the AB magnitude system from R-band-filter converted through {\sc SNAP}, since only these tow observations were done with the $R$ filter.
\label{table:sn13hx}
\end{table*}%

\begin{table*}
\caption{$g,r,i,z$ magnitudes of \br\/ and assigned errors in brackets.}
\begin{center}
\begin{tabular}{cccccccl}
\hline
\hline
Date & MJD & Phase*   &$g$ & $r$ & $i$ & $z$& Inst.\\
dd/mm/yy &  & (days) & & & &\\
\hline
16/01/15  & 57038.41 & -46.15  &     -    & $>$ 20.6   &     -     &    -      &CSS\\
16/02/15  & 57069.43 & -17.99  &   -      & 18.95$^{\dagger}$ (0.07)   &    -      &    -      &PSST\\
17/02/15  & 57070.43 & -16.99  &     -    & 18.86$^{\dagger}$  (0.07)   &    -      &     -     &PSST\\
26/02/15  & 57079.48 & -8.87  &    -     & 18.33  (0.07)   &     -     &     -     &CSS\\ 
10/03/15  & 57091.52 & 2.06 &     -    & 18.24 (0.07)   &    -      &      -    &CSS\\ 
11/03/15  & 57093.01 & 3.41 & 18.16 (0.01) & 18.25 (0.01) &  18.24 (0.01) & 18.40 (0.02)  &LT\\
14/03/15  & 57095.94 & 6.07 & 18.17 (0.01) & 18.21 (0.01) &  18.24 (0.02) & 18.40 (0.03)  &LT\\
19/03/15  & 57100.84 & 10.52 &  18.34 (0.07)   & 18.25 (0.07)    &  18.32 (0.07)    &    -      &LCO\\
19/03/15  & 57100.91 & 10.59 &  18.31 (0.07)   & 18.29 (0.10)    &  18.33 (0.07)    &    -      &LCO\\
21/03/15  & 57102.52 & 12.05 &      -   & 18.36 (0.07)    &      -    &     -     &CSS\\
23/03/15  & 57104.57 & 13.91 &  18.45 (0.07)   & 18.38 (0.07)    &  18.37 (0.07)    &      -    &LCO\\
24/03/15  & 57105.50 & 14.75 &     -    & 18.41$^{\dagger}$ (0.07)   &     -     &     -     &PSST\\
25/03/15  & 57106.51 & 15.67 &   -      & 18.44$^{\dagger}$ (0.07)   &     -     &     -     &PSST\\
27/03/15  & 57108.45 & 17.43 &  18.48 (0.07)   & 18.45 (0.08)    &  18.36 (0.07)    &     -     &LCO\\
31/03/15  & 57112.90 & 21.47 & 18.57 (0.05) & 18.40 (0.03) &  18.36 (0.04) & 18.45 (0.06)  &LT\\
04/04/15  & 57117.03 & 25.22 &  18.58 (0.08)   & 18.42 (0.07)    &  18.40 (0.11)    &    -      &LCO\\
07/04/15  & 57120.40 & 28.28 &  18.60 (0.09)   & 18.45 (0.08)    &  18.44 (0.11)    &     -     &LCO\\
10/04/15  & 57121.52 & 29.29 &    -     & 18.52 (0.07)    &    -      &     -     &CSS\\
10/04/15  & 57122.94 & 30.58 & 18.70 (0.01) & 18.55 (0.02) &  18.42 (0.01) & 18.48  (0.06)   &LT\\
11/04/15  & 57124.46 & 31.96 &  18.73 (0.08)   & 18.63 (0.08)    &  18.45 (0.11)    &     -     &LCO\\
12/04/15  & 57124.94 & 32.40 & 18.77 (0.01) & 18.61 (0.01) &  18.46 (0.02) & 18.50 (0.02)  &LT\\
14/04/15  & 57126.94 & 34.21 & 18.78 (0.01) & 18.61 (0.01) &  18.48 (0.02) & 18.48 (0.02)  &LT\\
16/04/15  & 57128.44 & 35.58 &     -    & 18.66$^{\dagger}$ (0.07)   &    -      &     -     &PSST\\ 
16/04/15  & 57128.52 & 35.65 &    -     & 18.69 (0.07)    &   -       &       -   &CSS\\
16/04/15  & 57128.99 & 36.07 &  18.96 (0.08)   & -    & -    &   -       &LCO\\
17/04/15  & 57129.88 & 36.88 &  18.97 (0.09)   & 18.70 (0.08)    &  18.49 (0.11)    &    -      &LCO\\
17/04/15  & 57129.93 & 36.93 & 18.95  (0.01)  & 18.66 (0.02) &  18.50 (0.01) & 18.49 (0.01)  &LT\\
18/04/15  & 57130.42 & 37.37 &     -   & 18.67 (0.07) &     -     &     -    &PSST\\  
20/04/15  & 57132.94 & 39.66 & 18.96  (0.02)  & 18.70 (0.01) &  18.52 (0.01) & 18.62 (0.02)  &LT\\
22/04/15  & 57134.79 & 41.34 &  19.03 (0.09)  & 18.74 (0.08) &  18.56 (0.11)    &     &LCO\\
23/04/15  & 57135.95 & 42.39 & 19.06  (0.02)  & 18.74 (0.02) &  18.48 (0.01) & 18.53 (0.01)  &LT\\
24/04/15  & 57136.52 & 42.91 &      -   & 18.71 (0.07)    &     -     &    -      &CSS\\
25/04/15  & 57138.93 & 45.10 & 19.06  (0.02) & 18.75 (0.02) &  18.49 (0.02) & 18.59 (0.02)  &LT\\
26/04/15  & 57139.39 & 45.52 &   19.10 (0.10)   & 18.79 (0.07)    &  18.60 (0.08)    &   -       &LCO\\
01/05/15  & 57144.29 & 49.96 &  19.21 (0.10)   & 18.79 (0.07)    &  18.58 (0.08)    &    -      &LCO\\
02/05/15  & 57144.88 & 50.50 & 19.23 (0.06) & 18.86 (0.07) &  18.56 (0.03) & 18.514 (0.03)  &LT\\
04/05/15  & 57147.47 & 52.85 &  19.24 (0.10)   & 18.94 (0.07)    &  18.78 (0.08)    &   -       &LCO\\
07/05/15  & 57149.52 & 54.71 &    -      & 18.92 (0.07)    &     -     &      -    &CSS\\
08/05/15  & 57150.93 & 55.99 & 19.25 (0.02) & 18.91 (0.01) &  18.80 (0.03) & 18.60 (0.02)  &LT\\
10/05/15  & 57152.24 & 57.18 &  19.24 (0.10)   & 18.97 (0.07)    &  18.79 (0.08)    &     -     &LCO\\
13/05/15  & 57155.52 & 60.16 &     -    & 18.97 (0.07)    &      -    &    -      &CSS\\ 
14/05/15  & 57156.42 & 60.97 &  19.25 (0.10)   & 19.05 (0.07)   &  18.68 (0.08)    &     -     &LCO\\
16/05/15  & 57158.86 & 63.19 & 19.23 (0.10) & 19.05 (0.04) &  18.74 (0.04) & 18.71 (0.02)  &LT\\
20/05/15  & 57162.87 & 66.83 & 19.25 (0.10) & 19.05 (0.05) &  18.72 (0.03) & 18.76 (0.05)  &LT\\
22/05/15  & 57164.52 & 68.33 &      -   & 19.10 (0.07)    &     -     &      -    &CSS\\ 
24/05/15  & 57166.91 & 70.50 & 19.45 (0.03) & 19.16 (0.03) &  18.80 (0.02) & 18.82 (0.02)  &LT\\
02/06/15  & 57175.31  & 78.12  & 19.57  (0.15)   & 19.28 (0.15)     & 18.91 (0.12)     &  -   &    LCO\\
03/06/15  & 57176.97 & 79.63 & 19.56 (0.08) & 19.31 (0.05) &  18.90 (0.04) & 19.03 (0.06)  &LT\\
11/06/15  & 57184.27 & 86.26 &  19.83 (0.10)   & 19.39 (0.20)    &  19.10 (0.12)    &    -      &LCO\\
12/06/15  & 57185.00 & 86.92 &      -    & 19.37 (0.20)    &   -       &      -    &CSS\\
20/06/15  & 57193.37 & 94.52 &-    &   - &  19.15 (0.20)    &     -     &LCO\\
30/06/15  & 57203.28 & 103.51 &  20.10 (0.20)   & 19.43 (0.15)    &  19.18 (0.10)    &      -    &LCO\\
02/07/15  & 57205.27 & 105.32 &  20.30 (0.30)   & 19.75 (0.20)    &  19.28 (0.20)    &     -     &LCO\\
08/12/15  & 57365.21 & 250.64 & & &      19.61 (0.08)   & &LT\\
13/12/15  & 57370.18 &  255.16   &  20.41 (0.06)  &  20.49 (0.07)   &  &    20.05 (0.07)   & LT\\
18/12/15  & 57375.15  &  259.67   & 20.44 (0.10) & 20.43 (0.10) & 19.63 (0.11) &  20.07 (0.11) & LT\\
17/01/16  & 57405.26  &  287.02  &  20.66 (0.10) &20.44 (0.09) & 19.82 (0.11) &  20.10 (0.15) & NTT \\
18/01/16  & 57406.04  &  287.73  &  20.52 (0.12) &20.49 (0.11) & 19.83 (0.12) &  20.13 (0.22)  & LT\\
25/01/16  & 57413.10  & 294.12   &  20.55 (0.14) &20.45 (0.12) & 19.88 (0.11) &  20.11 (0.21) & LT\\
05/02/16  & 57424.25  &  304.27  &  20.56 (0.10) &20.51 (0.11) &  19.90 (0.10)  & 20.19 (0.10)  & NTT \\
08/03/16  & 57456.24  &  333.32  &    20.56 (0.09) &20.51 (0.10) & 19.85 (0.15) &  20.47 (0.10)   & NTT \\
12/04/16  & 57491.11   & 365.00 &   20.91 (0.09) &20.74 (0.09)  &19.86 (0.15) &  20.54 (0.09)  & NTT \\

Host & &&&&&\\
01/05/10 - 01/10/14  &  &   & 22.39 (0.11) & 22.18 (0.12) & 21.70 (0.09) &22.15 (0.11)       &PS1\\
\hline
\end{tabular}
\end{center}
* Phase with respect to the r-band maximum.\\
${\dagger}$ $r$ in the AB magnitude system from $w$-band filter converted through {\sc SNAP}.
\label{table:sn15br}
\end{table*}%

\begin{table*}
\caption{$U,B,V,J,H,K$ magnitudes of \br\/ and assigned errors in brackets.}
\begin{center}
\begin{tabular}{ccccccl}
\hline
\hline
Date &MJD & Phase* &  $U$ &$ B$ &$V$ & Inst.\\
dd/mm/yy & & (days) & & & &\\
\hline
10/03/15  &  57092.07 & 2.56  &     -    &    -    &  18.00 (0.01)  &NTT\\  
11/03/15  &  57092.87 & 3.29 &  17.88 (0.10)   &18.44 (0.10)    &  18.04  (0.13)  &SWIFT\\  
12/03/15  &  57094.23 & 4.52 &      -   &     -   &  17.95 (0.01)  &NTT\\ 
13/03/15  &  57095.20 & 5.40  &      -   &    -    &  17.97 (0.02)  &NTT\\ 
14/03/15  &  57095.96 & 6.09  &	17.95 (0.13) &  18.48 (0.13)  &    18.05  (0.19)&  SWIFT\\  
19/03/15  &  57100.84 & 10.52  &     -    & 18.51 (0.10)   &  18.10 (0.07)     &LCO\\
19/03/15  &  57100.91 & 10.58  &     -    & 18.63 (0.08)   &  18.15 (0.07)     &LCO\\
20/03/15  &  57102.22 & 11.77  &	18.32 (0.11) &  18.64 (0.12)  &    18.18  (0.21)&  SWIFT\\
23/03/15  &  57104.57 & 13.91  &      -   & 18.67 (0.08)   &  18.21 (0.07)     &LCO\\
23/03/15  &  57105.14 & 14.42  &	18.65 (0.15) &  18.73 (0.19)  &    18.21  (0.21)&  SWIFT\\
26/03/15  &  57107.93 & 16.96  &	18.80 (0.15) &  18.83 (0.19)  &    18.30  (0.23)&  SWIFT\\
27/03/15  &  57108.45 & 17.43  &     -    & 18.82 (0.08)   &  18.34 (0.07)     &LCO\\
27/03/15  &  57109.30 & 18.20  &    -     &    -    &  18.35 (0.06)  &NTT\\ 
28/03/15  &  57110.04 & 18.87  &   -      &   -     &  18.30 (0.01)  &NTT\\ 
30/03/15  &  57112.15 & 20.79 &           19.02 (0.18) &  19.01 (0.13)  &    18.38  (0.21)&  SWIFT\\
04/04/15  &  57116.99 & 25.18  &           19.22 (0.18) &  19.04 (0.18)  &    18.43  (0.21)&  SWIFT\\
04/04/15  &  57117.03 & 25.22  &  -       & 19.05 (0.15) &  18.44 (0.10)     &LCO\\
07/04/15  &  57119.98 & 27.90  &	       19.42 (0.18) &  19.07 (0.18)  &    18.50  (0.21)&  SWIFT\\
07/04/15  &  57120.40 & 28.28  &    -     & 19.08 (0.15)   &  18.49 (0.14)     &LCO\\
11/04/15  &  57124.46 & 31.96  &    -     & 19.17 (0.11)   &  18.58 (0.09)     &LCO\\
16/04/15  &  57128.99 & 36.07  &    -     & 19.19 (0.13)   &  18.76 (0.10)     &LCO\\
17/04/15  &  57129.88 & 36.88  &    -     & 19.30 (0.11)   &  18.79 (0.11)     &LCO\\
18/04/15  &  57130.81 & 37.73  &    -     & 19.45 (0.11)   &  18.80 (0.09)     &LCO\\
22/04/15  &  57134.79 & 41.34  &    -     & 19.47 (0.20)   &  18.85 (0.11)     &LCO\\
25/04/15  &  57138.13 & 44.37 &      -   &   -     &  18.84 (0.03)  &NTT\\
26/04/15  &  57139.39 & 45.52 &      -   & 19.44 (0.12)   &  18.81 (0.10)     &LCO\\
01/05/15  &  57144.29 & 49.96  &     -    & 19.43 (0.12)   &  18.84 (0.10)     &LCO\\
04/05/15  &  57147.47 & 52.85  &     -    & 19.42 (0.12)   &  18.84 (0.10)     &LCO\\
10/05/15  &  57152.24 & 57.18  &     -    & 19.59 (0.12)   &  18.86 (0.10)     &LCO\\
14/05/15  &  57156.42 & 60.97  &     -    & 19.67 (0.20)   &  19.15 (0.10)     &LCO\\
\hline
\hline
Date & MJD & Phase* &  $J$ & $H$ & $K$ &  Telescope  \\
dd/mm/yy  & (days)  &   &\\
\hline
11/03/15  & 57093.22 & 3.60 & 17.80  (0.02)    &  17.48  (0.03)	 &   17.27  (0.06)  & NTT\\
16/04/15  &  57129.14 &36.21	 &  17.76  (0.01)    &  17.23  (0.01)	   & 17.01  (0.02)  & NTT\\
16/01/16  & 57404.25  & 286.10 & 19.06  (0.05)   &  18.95  (0.05) &	   17.61  (0.08)   &   NTT\\
06/02/16  & 57425.16  & 305.09 & 19.293  (0.12)   &  19.105  (0.10)  &   17.630  (0.08)       &   NTT\\
07/03/16  & 57455.22  & 332.40 & 19.098  (0.25)   &  19.239  (0.08)  &   17.537  (0.12)      &   NTT\\
13/04/16  & 57492.09  & 365.89 & 19.425  (0.12) &    19.514  (0.18)  &   17.659  (0.08)      &   NTT\\
\hline
\end{tabular}
\end{center}
* Phase with respect to the r-band maximum.\\
\label{table:sn15br2}
\end{table*}%

\begin{table*}
\caption{{\it Swift}+UVOT $uvw2,uvm2,uvw1$ magnitudes of \hx\/ and \br\/ and assigned errors in brackets.}
\begin{center}
\begin{tabular}{cccccc}
\hline
\hline
Date & MJD & Phase*   &$uvw2$ & $uvm2$ & $uvw1$ \\
dd/mm/yy &  & (days)  &  & &\\
\hline
\multicolumn{6}{c}{\hx}\\
\hline
21/02/14&	56709.88& 22.56 &	17.67  .10 &17.44  .08&	17.17  .09\\
24/02/14& 	56712.82& 25.17&	17.87  .10	& 17.75  .08&	17.46  .09\\
27/02/14& 	56715.36& 27.43&	18.10  .12	& 18.05  .16&	17.63  .09\\
03/03/14& 	56719.16& 30.81&	18.49  .12	& 18.13  .10&	18.01  .11\\
05/03/14& 	56721.17& 32.59&	18.58  .13	& 18.38  .11&	18.07  .13 \\
08/03/14& 	56724.56& 35.61&	18.85  .13	& 18.58  .11&	18.34  .12 \\
27/04/14& 	56774.90& 80.35&	- & $>$21.0 &-	 \\
28/04/14& 	56775.03& 80.47&	 -	& -&	$>$21.8 \\
30/04/14& 	56777.36& 82.54&	$>$22.0	& -&- 	\\
07/07/14& 	56845.79& 143.37&	$>$21.9	& -&	 -	\\
21/07/14& 	56859.07& 155.17 &		-& -&	$>$21.6   \\
24/07/14& 	56862.46& 158.19& 	-	& $>$21.1 & -	\\
\hline

\multicolumn{6}{c}{\br\/}\\
\hline
11/03/15 &	57092.87 & 3.29  &	18.70  (0.11)&	18.67  (0.17)	&18.25  (0.11)\\
14/03/15 &	57095.96 	&6.09  &19.01  (0.17)	&18.87  (0.22)&	18.50  (0.18) \\
20/03/15 &	57102.22 	&11.77 &19.37  (0.14)	&19.24  (0.19)&	18.83  (0.15)\\
23/03/15 &	57105.14 &14.42 &	19.57  (0.21)&	19.44  (0.32)	&18.85  (0.15) \\
26/03/15 &	57107.93 &16.96 &	19.78  (0.18)	&19.13  (0.19)&	18.92  (0.16) \\\
30/03/15 &	57112.15 & 20.79 &	20.06  (0.19)&	19.77  (0.24)&	19.29  (0.18) \\
04/04/15 &	57116.99 & 25.18  &	20.26  (0.30)&	19.94  (0.26)&	19.47  (0.24) \\
07/04/15 &	57119.98& 27.90  & 	20.39  (0.24)&	20.03  (0.28)&	19.62  (0.28) \\
11/11/15 &   57337.64  &   225.60 & 21.15  (0.23) & -  & \\
15/11/15 &   57341.03  &   228.68 & 20.92  (0.24) &  -  & \\
21/01/16 &   57407.80  &   289.33 & -  &  -  & 20.98  (0.38)  \\
25/01/16 &   57412.04  &   293.18 &  21.03  (0.27) & 21.04  (0.30)  &20.65  (0.25) \\
\hline
\end{tabular}
\end{center}
* Phase with respect to the r-band maximum.\\
\label{table:sn15bruv}
\end{table*}%

\begin{table*}
\caption{Magnitudes in $B,V,g,r,i$ of the local sequence stars in the field of \hx.}
\begin{center}
\begin{tabular}{lccccc}
\hline
\hline
Star & $B$ & $V$ & $g$ & $r$ & $i$ \\
\hline
1 &21.06 (0.09)& 19.24 (0.05) & 19.35 (0.03)  & 17.81 (0.03) & 16.55 (0.02)\\
2 &18.96 (0.06)& 16.68  (0.02) & 16.83 (0.02) & 15.26  (0.03) & 13.81 (0.02)\\
3 &19.59  (0.07)& 18.17  (0.03) & 17.98 (0.02)  & 17.30 (0.02)& 16.69 (0.02)\\
4 & 	-   & 22.89  (0.09) & 22.90 (0.08) & 21.17 (0.08)& 20.65 (0.05)\\
5 &20.56 (0.08)& 18.76   (0.05) & 18.90 (0.04) & 17.39 (0.05) & 15.94 (0.02)\\
6 &19.29 (0.07)& 17.95 (0.06)  & 17.72 (0.03) & 17.20 (0.03)& 16.65 (0.03)\\
7 & -	   & 22.50  (0.09) & 22.17 (0.06) & 20.41  (0.05) & 18.40 (0.04)\\
8 & 	-   & 22.50 (0.08)  & 22.61 (0.09) & 21.82 (0.08)& 21.34 (0.06)\\
9 & 	-   & 22.54  (0.08) & 22.77 (0.09) & 21.11 (0.07) & 19.13 (0.04)\\
10 &17.42 (0.05)& 16.04 (0.02)  & 15.83 (0.02) & 15.15 (0.02) & 14.67 (0.03)\\
 11 &  -    & 22.22 (0.08)  & 22.38 (0.08) & 21.07 (0.07) & 20.21 (0.05)\\
12 &   -    & 22.36 (0.08)  & 22.50 (0.09) & 21.15 (0.07) & 20.28 (0.05)\\
\hline
\end{tabular}
\end{center}
\label{table:ssj13hx}
\end{table*}%

\begin{table*}
\caption{Magnitudes in $B,V,g,r,i,z$ of the local sequence stars in the field of \br.}
\begin{center}
\begin{tabular}{lcccccc}
\hline
\hline
Star &$B$ & $V$& $g$ & $r$ & $i$ & $z$ \\
\hline
1 &16.50 (0.02)&16.02 (0.02)&16.19 (0.01)& 15.96 (0.01) & 15.83 (0.02)   & 15.83 (0.01)\\
2 &18.02 (0.02)&17.47 (0.02)&17.67 (0.01)& 17.38  (0.01) & 17.22 (0.01)  & 17.18 (0.01)\\
3 &20.57 (0.05) & 19.32 (0.03) &19.86 (0.02)& 18.85  (0.01) & 18.27 (0.01)  & 17.97 (0.01)\\
4 &17.41 (0.02) &16.51 (0.02)&16.91 (0.01) & 16.16  (0.01) & 15.85 (0.01) &  15.69 (0.01)\\
5 &19.03 (0.03)& 17.89 (0.02) &18.41 (0.01)& 17.44   (0.01) & 16.97 (0.01)  & 16.72 (0.01)\\
6 &19.12 (0.03) & 18.65 (0.03) &18.81 (0.02)& 18.60 (0.01)  & 18.49 (0.01) & 18.47 (0.01)\\
7 &18.33 (0.03) & 17.80 (0.02)& 17.99 (0.01)  & 17.70  (0.01) & 17.55 (0.01) & 17.52 (0.01)\\
8 &20.17 (0.05) & 18.51 (0.03)& 19.07	(0.02)  & 17.89 (0.01)  & 16.96 (0.01) & 16.49 (0.01)\\
9 &18.07 (0.02) &17.44 (0.02)& 17.68 (0.01) & 17.29  (0.01) & 17.08 (0.01) & 17.01 (0.01)\\
10 &20.75 (0.05) &19.59 (0.05) &20.09 (0.02)& 19.15 (0.01)  & 18.63 (0.01) & 18.37 (0.01)\\
11 &21.48 (0.06) & 20.15 (0.05)& 20.72 (0.03)   & 19.63 (0.02)  & 19.03 (0.02) & 18.66 (0.01)\\
\hline
\end{tabular}
\end{center}
\label{table:ssj15br}
\end{table*}%

\begin{table*}
\caption{Journal of spectroscopic observations.}
\begin{center}
\begin{tabular}{cccccl}
\hline
\hline
Date & MJD & Phase* &Range$^{\dagger}$ & Resolution & Instrumental  \\
dd/mm/yy &  & (days)  &(\AA)  & (\AA) &Configuration\\
\hline
\multicolumn{6}{c}{\hx}\\
\hline
31/01/14 &   56688.43   & 3.49 &3100--6900 & 1.5/2.5  & ANU 2.3m+WiFeS+B300/R300\\    
19/02/14 &   56708.03    & 20.82 &3600--9300 & 18 & NTT+EFOSC2+gm13\\                   
20/02/14 &   56709.02      & 21.70 &3600--9300 & 18 & NTT+EFOSC2+gm13\\ 
21/02/14 &   56710.02      & 22.58  & 6000--10000  & 16 & NTT+EFOSC2+gm16\\      
22/02/14 &   56711.02      & 23.47  & 3300--7500  & 13 & NTT+EFOSC2+gm11\\      
24/02/14 &   56712.44    & 24.84 & 3400--6900& 1.5/2.5   & ANU 2.3m+WiFeS+B300/R300\\   
28/02/14 &   56717.04      & 28.80  & 3300--10000  & 13/16 & NTT+EFOSC2+gm11/gm16\\  
04/03/14 &   56720.42   & 31.93 & 3400-6900 &1.5/2.5   & ANU 2.3m+WiFeS+B300/R300\\   
08/03/14 &   56725.03      &  35.87 & 6000--10000  & 16 & NTT+EFOSC2+gm16\\   
09/03/14 &   56726.06     & 36.78  & 3300--10000  & 13/16 & NTT+EFOSC2+gm11/gm16\\ 
10/03/14 &   56727.02      & 37.62  &  3300--7500  & 13 & NTT+EFOSC2+gm11\\   
14/03/14 &   56730.39   & 40.79  & 3400--6900 &1.5/2.5   & ANU 2.3m+WiFeS+B300/R300\\   
21/09/14 &   56922.10     & 210.27  &  6900--7700  & 18 & NTT+EFOSC2+gm13\\   
22/10/14 &   56953.07     & 237.67  & 4500--9300  & 11 & VLT+FORS2+GRIS\verb|_|300V\\ 
\hline
\multicolumn{6}{c}{\br}\\
\hline
10/03/15 & 57092.07 & 2.56 &3600--9300& 18 &  NTT+EFOSC2+gm13\\ 
12/03/15 & 57093.56 & 3.91 & 4100--6900 & 1.5/2.5 & ANU 2.3m+WiFeS+B300/R300\\   
12/03/15 & 57094.23 & 4.52 &3300--7500 & 13 &  NTT+EFOSC2+gm11\\ 
13/03/15 & 57095.20 & 5.40 &6000--10000 & 16  &  NTT+EFOSC2+gm16\\ 
27/03/15 & 57109.30 & 18.20 &3300--10000 & 18  &  NTT+EFOSC2+gm13\\ 
28/03/15 & 57100.05 & 18.88 &3600--9300& 18 &  NTT+EFOSC2+gm13\\ 
12/04/15 & 57125.06 & 32.51 &3600--9300& 18 &  NTT+EFOSC2+gm13\\ 
25/04/15 & 57138.13 & 44.37 &3600--9300& 18 &  NTT+EFOSC2+gm13\\ 
29/06/15 & 57202.13 & 102.47 &3600--9300& 4.3/5.8 &  UH 2.2m+SNIFS+B/R\\
08/01/16 & 57396.24 & 278.82 &3600--9300& 18 &  NTT+EFOSC2+gm13\\ 
12/03/16 &   57460.26    & 336.98  & 2811--24511  & 1.0/1.1/3.3 & VLT+XSHOOTER+UV/OPT/NIR\\ 
09/04/16 &   57488.08    & 362.24 & 2811--24511  & 1.0/1.1/3.3 & VLT+XSHOOTER+UV/OPT/NIR\\ 
10/04/16 & 57489.09& 363.16 &3600--9300& 18 &  NTT+EFOSC2+gm13\\ 
\hline
\end{tabular}
\end{center}
* Phase with respect to the r-band maximum.\\
$\dagger$ Range with respect to the observed frame.
\label{table:sp}
\end{table*}%

\section{Sequence stars and additional plots}\label{sec:ss}

\begin{figure*}
\center
\includegraphics[width=8cm,height=7.93cm]{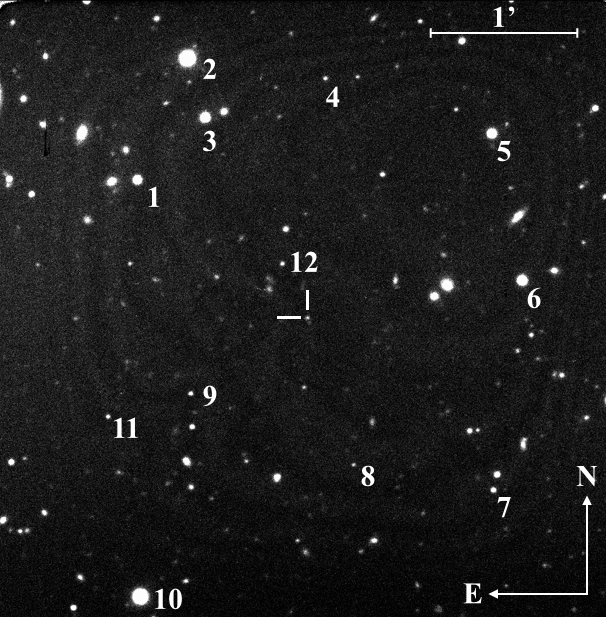}
\includegraphics[width=8cm]{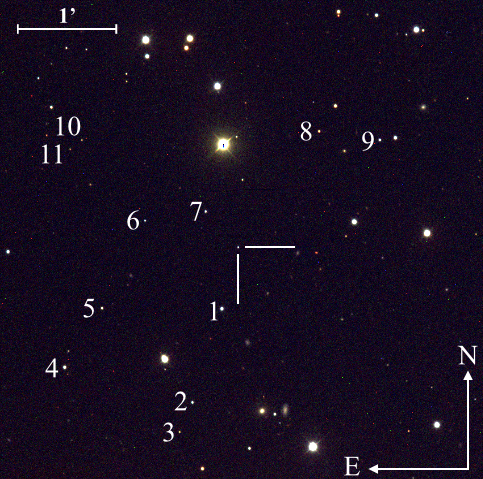}
\caption{Left: NTT+EFOSC2+$r$ image of \hx. The SN position is indicated with cross white marks. The sequence of stars in the field used to calibrate the optical and NIR magnitudes of \hx\/ is indicated. Right: LT+IO+$g$/$r$/$i$ image of \br\/ (cross white marks). The sequence of stars in the field used to calibrate the optical and NIR magnitudes of \br\/ is indicated}
\label{fig:fc}
\end{figure*}

\bsp	
\label{lastpage}
\end{document}